\documentclass[aps,pre,onecolumn,floatfix,superscriptaddress]{revtex4}

\usepackage{bm}
\usepackage{graphicx}
\usepackage{amssymb,amsfonts,amsmath}
\usepackage{epsf}
\usepackage{subfigure}
\usepackage{epstopdf}
\allowdisplaybreaks
\usepackage{mathrsfs}
\usepackage{hyperref}
\usepackage{upgreek}
\usepackage{xcolor}
\usepackage{soul}

\newcommand{\be}{\begin{equation}}
\newcommand{\ee}{\end{equation}}
\newcommand{\bea}{\begin{eqnarray}}
\newcommand{\eea}{\end{eqnarray}}

\allowdisplaybreaks

\begin{document}

\title{\bf Dynamic structure factor of a monatomic cubic crystal}

\author{Arsène Yerle}
\affiliation{Laboratoire Matière et Systèmes Complexes, Université Paris Cité and CNRS (UMR 7057), 10 rue Alice Domon et Léonie Duquet, 75013 Paris, France}
\author{Pierre Gaspard}
\email{Gaspard.Pierre@ulb.be; \vfill\break ORCID: 0000-0003-3804-2110.}
\affiliation{Center for Nonlinear Phenomena and Complex Systems, Universit{\'e} Libre de Bruxelles (U.L.B.), Code Postal 231, Campus Plaine, B-1050 Brussels, Belgium}
\author{Jo\"el Mabillard}
\email{Joel.Mabillard@u-paris.fr; \vfill\break ORCID: 0000-0001-6810-3709.}
\affiliation{Laboratoire Matière et Systèmes Complexes, Université Paris Cité and CNRS (UMR 7057), 10 rue Alice Domon et Léonie Duquet, 75013 Paris, France}

\begin{abstract}
The spectral function of density fluctuations, also known as the dynamic structure factor, of a monatomic cubic crystal with vacancies is derived from the macroscopic equations describing transport in crystalline solids. The resonances of the spectral function are identified as  a Brillouin doublet of sound propagation, a central Rayleigh peak of heat diffusion, as for perfect crystals,  and another central sharp peak associated with vacancy diffusion. Analytical expressions for the heat and vacancy diffusivities,  speeds of sound, and sound damping coefficients are obtained. The theoretical results are compared to molecular dynamics simulations of a face-centered cubic crystal of hard spheres.
\end{abstract}

\maketitle

%%%%%%%%%%%%%%%%%%%%%%%%%%%%%%%%%%%%%%%%%%%%%%%%%
\section{Introduction}
\label{sec:intro}
%%%%%%%%%%%%%%%%%%%%%%%%%%%%%%%%%%%%%%%%%%%%%%%%%
In dynamic scattering experiments a beam of neutrons~\cite{vH54} or light~\cite{BP76,BY80} is directed at a sample, and the fluctuating scattered signal intensity is measured. The molecular theory of scattering states that the power spectrum derived from the time-correlation function of the scattered beam intensity is related to the spectrum of the medium density fluctuations, also referred to as the dynamic structure factor (DSF). On the theoretical side, the DSF is obtained from the dissipative hydrodynamic equations, which describe transport phenomena in the medium.  The resonances or peaks of the DSF correspond to the dispersion relations of the slow modes of the sample. In this way, the resonances of the spectrum are  associated with transport phenomena, for example, the diffusion of heat or the propagation and damping of sound in a simple fluid~\cite{M66,F75,BP76,BY80}. The DSF is thus a crucial object to study the transport properties of a given system and make the connection from microscopic dynamics to experiments.

Macroscopic transport equations stem from the slow modes of the system. These modes are characterized by frequencies vanishing with their wave number, implying that in the hydrodynamic limit (where the wave number is small), the relaxation to equilibrium is realized on macroscopic time scales~\cite{LL80a}. Consequently, these modes are considered slow and correspond to  transport phenomena on macroscopic spatiotemporal scales. The symmetries of the system dictate the number of modes. Each fundamental conservation law, like those for mass, momentum, and energy, introduces a corresponding slow mode~\cite{GM84}, expressed as the mass density $\rho({\bf r}, t)$, momentum density $g^a({\bf r}, t)=\rho  v^a({\bf r}, t)$, where $ v^a({\bf r}, t)$ is the velocity field, and energy density $\epsilon({\bf r}, t)$, respectively, at any time $t$ and point ${\bf r}$ of the system. These five modes are characteristic of a one-component fluid and are also found in monatomic crystals.  Additionally, each broken continuous symmetry gives rise to a Nambu-Goldstone mode~\cite{F75}. In a crystal, the continuous translational symmetry is broken into one of 230 discrete space groups defining the crystalline structure, adding three modes, identified as the displacement field $u^a({\bf r}, t)$~\cite{MPP72,FC76}, to the five modes stemming from the fundamental conservation laws. In this work, we restrict to the space groups with cubic symmetry.

The eight equations describing transport in monatomic crystals were obtained in the early 1970s~\cite{MPP72,FC76}, with a phenomenological approach. The hydrodynamic modes were identified as  three pairs of sound modes propagating in the longitudinal and transverse directions, a diffusive heat mode, and an eighth mode conjectured to be the mode of vacancy diffusion. In a real crystal, there are vacancies, i.e. lattice sites that are not occupied by any particle. Particles next to a vacancy can move into these empty sites. As particles keep moving into nearby vacant sites, the vacancy itself appears to diffuse through the crystal until it eventually reaches the surface. From the late 1990s, the transport equations are derived from the underlying microscopic dynamics with methods of statistical mechanics~\cite{SE93,S97,MG20,MG21,H22,H23}. Recently, it was shown in Ref.~\cite{MG25} that for a proper characterization of the diffusion of vacancies, and the identification of  corresponding Fick's law, the transport equations should be expressed in terms of the chemical potential of vacancies $\mu_{y}$ and the molar concentration of vacancies $y$. In this way, the eighth hydrodynamic mode of nonperfect monatomic crystals is shown to be indeed a mode of vacancy diffusion.

While the theoretical expression of the DSF for a one-component fluid is well established since the 1960s~\cite{M66} (based on earlier works of Landau and Placzek~\cite{LL84}) and the 1970s~\cite{BP76,BY80,F75} (from the theory of hydrodynamic fluctuations~\cite{LL80b}),  the case of crystalline solids has been treated recently~\cite{MG24b} but only for perfect crystals. In this case, there are no vacancies; every site of the lattice is occupied by exactly one particle. The perfect crystal has thus seven dispersion relations, and the DSF resonances correspond to the propagating sound modes and the heat diffusion mode. In this work, we generalize the results of Ref.~\cite{MG24b}, by going from idealized perfect crystals to the more realistic case of crystals with vacancies. The first goal is to find and  highlight the additional central resonance in the DSF  that is anticipated as a result of vacancy diffusion. The presence of this resonance provides further evidence for the identification of the eighth hydrodynamic mode. The other objective of this work is to obtain analytical expressions for the speeds of sound, the sound damping coefficients, and the heat and vacancy diffusivities, including the explicit contributions stemming from the presence of vacancies.

This work is organized as follows. In Sec.~\ref{sec:hydro}, the eight equations describing transport in cubic crystalline solids are introduced. The system is then closed using crystal thermodynamics.  By choosing special directions of the wave vector~${\bf q}$, the closed system is decoupled into a set of four longitudinal equations and two sets of two transverse equations. In  Sec.~\ref{sec:DSF}, we first obtain an exact analytical expression for the dynamic structure factor. From the poles of the spectral function, the dispersion relations are identified, including the one associated with the diffusion of vacancies. Analytical expressions for the heat and vacancy diffusivities,  speeds of sound, and sound damping coefficients are obtained. With an expansion in the limit where the wave number $q$ is small, the intermediate scattering function (ISF) is derived, as well as an approximate expression for the DSF. In this way, the resonances of the DSF are identified as Lorentzian functions. In Sec.~\ref{sec:MDHS}, the theoretical results  are applied to a face-centered cubic (fcc) hard-sphere crystal with vacancies. Concluding remarks and perspectives are presented in Sec.~\ref{sec:conclusion}. Further details on the derivations and on the simulation are given in the appendices.

{\it Notations and conventions.} The spatial coordinates are labeled by the Latin indices $a, b, c, \ldots = x, y, z$, and the particles or vacancies by the indices $i,j,\ldots = 1,2,\ldots$ . Unless explicitly stated, Einstein's convention of summation over repeated indices is adopted. $k_{\rm B}$ denotes Boltzmann's constant and ${\rm i}\equiv\sqrt{-1}$. The notation $\delta \hat X \equiv \hat X- \langle \hat X\rangle_{\rm eq}$ defines the fluctuation of the quantity $\hat X$ around equilibrium. The hat $\hat{}$ represents a microscopic quantity, and the angular brackets $\langle \cdot \rangle_{\rm eq}$ stand for the ensemble average with respect to the equilibrium grand canonical distribution.  The superscript $^{\top}$ denotes matrix transpose.  Cubic symmetric rank-four tensors $T^{abcd}$ are expressed   in terms of three independent components $T_{11}=(T^{xxxx}+T^{yyyy}+T^{zzzz})/3$, $T_{12}=(T^{xxyy}+T^{yyzz}+T^{zzxx})/3$ and $T_{44}=(T^{xyxy}+T^{yzyz}+T^{zxzx})/3$ using Voigt notation. Given a function ${f}({\bf r}, t)$, we denote the Fourier-Laplace transform as $\tilde{f}({\bf q}, z)$ and  the Fourier transform as ${f}({\bf q}, t)$. Dimensionless quantities in the results of the numerical simulations are denoted by an asterisk.

%%%%%%%%%%%%%%%%%%%%%%%%%%%%%%%%%%%%%%%%%%%%%%%%%
\section{Hydrodynamic equations of the monatomic cubic crystal}
\label{sec:hydro}
%%%%%%%%%%%%%%%%%%%%%%%%%%%%%%%%%%%%%%%%%%%%%%%%%

The dynamic structure factor is defined as the Fourier transform of the time-dependent correlation function of density fluctuations. To determine the time evolution of the density fluctuations, which are governed by a macroscopic equation coupled with the other  transport equations of the crystal, we close the system of equations using thermodynamic relations. Then, by choosing special directions of the wave vector, we decompose the system into independent sets of longitudinal and transverse equations.

We consider a three-dimensional crystal with a cubic-symmetric lattice. The crystalline lattice has $N_0$ sites, and the averaged lattice density is $n_0\equiv N_0/V$ where $V$ is the volume of the crystal. On the lattice, there are $N$ identical particles of mass $m$, and the averaged particle density is $n_{\rm eq} \equiv N/V$. The particle density $n({\bf r},t)$ is a function of space and time and satisfies $N=\int_V\text{d}{\bf r}\ n({\bf r},t)$ at any time. The mass density satisfies $\rho({\bf r},t) = mn({\bf r},t)$ and the equilibrium density $\rho_{\rm eq}= mn_{\rm eq}$. Moreover,  the averaged mass density of a perfect crystal is given by $\rho_{0}\equiv mn_{0}$. The number of vacancies $N_{\rm v}\equiv N_0-N$ is conserved if there is no creation of vacancy-interstitial pairs, which we assume to be the case here. The density of vacancies $n_{\rm v}({\bf r},t)$ satisfies $N_{\rm v}=\int_V{\rm d}{\bf r}\ n_{\rm v}({\bf r},t)$ at any time, and the averaged density of vacancies is $n_{\rm v, eq}\equiv N_{\rm v}/V$. We also define the molar fraction of vacancies $y({\bf r},t)\equiv n_{\rm v}({\bf r},t)/n_{\rm 0}$, which plays a central role in the following.

\subsection{Transport equations}
The macroscopic evolution of the eight slow modes of the monatomic cubic crystal is given by transport equations for the densities $(\rho,\epsilon_0,\rho v^a, u^a)$, where $\epsilon_0$ is the energy density in a frame where ${\bf v}=0$. Linearized around the equilibrium rest state, these eight equations, given in Ref.~\cite{MG25},  read
\begin{align}
 \partial_t\,  \delta \rho & =- \rho \, \nabla^a \delta v^a\;,\label{eq:drho}\\
\partial_t \, \delta{\mathfrak \epsilon}_0 &=-(\epsilon_0+p) \nabla^a \delta v^a + \kappa^{\prime} \, \nabla^2 \delta T + \xi \, \nabla^a \nabla^b \delta \phi^{ab} - \xi \, \mu_{y,\boldsymbol{\phi}} \, \nabla^2\delta\phi^{aa} - {\cal K}_y \, \nabla^2 \delta y \, ,\label{eq:de} \\
 \rho \, \partial_t \, \delta v^b & = \nabla^a \delta \sigma^{ab}  + \eta^{abcd}\, \nabla^a\nabla^c \delta v^d \, , \label{eq:drhov}\\
 \partial_t \, \delta u^b & = \delta v^b +\frac{\xi^{\prime}}{T} \, \nabla^b \delta T + \zeta \, \nabla^a \delta \phi^{ab} - \zeta \, \mu_{y,\boldsymbol{\phi}} \, \nabla^b\delta \phi^{aa}- {\cal D}_{y} \, \nabla^b\delta y \, , \label{eq:du}
\end{align}
where $p$ is the hydrostatic pressure, $T$ the temperature field, $\sigma^{ab}\equiv-p\delta^{ab}+\phi^{ab}$ the stress tensor, $\phi^{ab}$ the excess stress tensor, $\mu_{y,\boldsymbol{\phi}} \equiv \frac{1}{3}\, {\rm tr}\left(\frac{\partial\mu_y}{\partial\boldsymbol{\phi}}\right)_{T,y}$, and
\begin{align}
{\cal D}_y \equiv \left(\frac{\partial\mu_y}{\partial y}\right)_{T,\boldsymbol{\phi}} \zeta\;, &&
{\cal K}_y &\equiv \left(\frac{\partial\mu_y}{\partial y}\right)_{T,\boldsymbol{\phi}} \xi \, , &&
\kappa^{\prime} \equiv \kappa - T \left[\frac{\partial(\mu_y/T)}{\partial T}\right]_{y,\boldsymbol{\phi}} \xi \, ,&&
\xi^{\prime} \equiv \xi - T^2 \left[\frac{\partial(\mu_y/T)}{\partial T}\right]_{y,\boldsymbol{\phi}} \zeta \, . \label{kappa-prime,K_y,xi-prime}
\end{align}
The first term on the right-hand side of each transport equation~\eqref{eq:drho}-\eqref{eq:du} is known as the Euler or conservative term. If all other contributions were disregarded, these terms would simply lead to conservative and reversible equations. The remaining terms are diffusive and irreversible because they contribute to the production of entropy and they break time-reversal symmetry~\cite{MG20,MG21,MG25}. These irreversible terms appear with transport coefficients and are associated with transport phenomena: $\kappa$ is the heat conductivity governing heat diffusion, $\eta^{abcd}$ is the viscosity tensor, linked to momentum transport, and $\zeta$ is the vacancy conductivity, which  leads to vacancy diffusion. In addition, $\xi$ is the vacancy thermodiffusion coefficient coupling heat and vacancy transports. There are no rank-three tensors in the transport equations, as a consequence of the cubic symmetry, in agreement with Curie’s principle~\cite{C1894}.  Microscopic formulae for all the transport coefficients exist as Green-Kubo or Einstein-Helfand formulae~\cite{S97,MG20,MG21,MG25}. The transport coefficients $\kappa, \zeta, \xi$  and the three independent components of the viscosity tensor, $\eta_{11}, \eta_{12}$ and $\eta_{44}$ (using Voigt notation for cubic symmetry),  are computed for a hard-sphere cubic crystal in Refs.~\cite{MG24a,MG25}.

The molar fraction of vacancies $y$ is a function of the density $\rho$ and the strain tensor $u^{ab}\equiv(\nabla^au^b+\nabla^bu^a)/2$. Indeed, under nonequilibrium conditions, the presence of vacancies or a strain of the lattice changes the local mass density $\rho$ according to the relation 
\begin{align}
	\rho &= \rho_{0} \left( 1 - y - u^{aa} \right)\;,\label{eq:relrhoyuaa}
\end{align}
where $\rho_{0}= mn_{0}$ is the product of the   mass $m$ of the individual particles and the averaged density of lattice sites $n_{\rm 0}$. With the transport equations~\eqref{eq:drho} and \eqref{eq:du}, the time evolution equation of the molar fraction of vacancies is given by
\begin{align}
 \partial_t \, \delta y = -\frac{\xi^{\prime}}{T} \, \nabla^2\delta  T - \zeta \, \nabla^a \nabla^b \delta \phi^{ab} + \zeta \, \mu_{y,\boldsymbol{\phi}} \, \nabla^2 \delta \phi^{aa} + {\cal D}_y \, \nabla^2\delta  y \, ,\label{eq:dy}
\end{align}
where the last term expresses Fick’s law for vacancies, leading to the identification of $ {\cal D}_y$ as the vacancy diffusion coefficient~\cite{MG25}. We note that the coefficients $\zeta$, $\xi$, and $\xi^\prime$ are of order $\mathcal{O}(y)$ and, therefore, they vanish with the molar fraction of vacancies in the limit $y\to 0$ of a perfect crystal, as shown in Ref.~\cite{MG25}.

\subsection{Closed system of equations and crystal thermodynamics}

To solve for the time evolution of density fluctuations, a closed system of equations is first obtained using crystal thermodynamics~\cite{W98}. We select the set of variables $(\delta y, \delta T, \delta v^a, \delta u^{ab})$ because, unlike the energy density, the variables $(\delta T, \delta v^a,  \delta u^{ab})$  are statistically independent. This independence ensures their equal-time cross-correlation functions vanish, a property also found in fluids and perfect crystals. Due to the relation~\eqref{eq:relrhoyuaa}, however, the variable $\delta y$ is not statistically independent from $\delta u^{ab}$ (and $\delta \rho$). The set of closed equations, derived in Appendix~\ref{app:TH}, reads
\begin{align}
\partial_t   \delta y& = {\cal D}_y\nabla^2 \delta y -\frac{\xi'}{T}\nabla^2  \delta T- G_T^{abcd}  \zeta  \nabla^a \nabla^b  \delta u^{cd}\;,\label{eq:deltay}\\
 \frac{c_v}{T} \partial_t\delta T + \varsigma_y  \partial_t\delta y+\frac{\alpha B_T}{\rho} \partial_t\delta u^{aa} &=- \frac{{\cal K}_y}{\rho T}  \nabla^2 \delta y+ \frac{ \kappa^{\prime} }{\rho T}\nabla^2 \delta T+ \frac{G_T^{abcd}  \xi}{\rho T} \nabla^a \nabla^b  \delta u^{cd}\;,\label{eq:deltaT}\\
\partial_t  \delta v^a& = - \frac{\pi_y}{\rho} \nabla^a \delta y- \frac{\alpha B_T }{\rho } \nabla^a \delta T + \frac{\eta^{abcd}}{\rho} \nabla^b\nabla^c \delta v^d + \frac{B_T^{abcd}}{\rho}\nabla^b \delta u^{cd}\;,\label{eq:deltav}\\
\partial_t  \delta u^{ab}& =-{\cal D}_y\nabla^a\nabla^b \delta y+\frac{ \xi^{\prime}}{T}\nabla^a\nabla^b\delta T+ \frac{1}{2}\left( \nabla^a\delta v^b+\nabla^b\delta v^a\right) \notag\\
&+\frac{ \zeta}{2}\left(G_T^{adef}\delta^{bc}+G_T^{bdef}\delta^{ac} \right) \nabla^c\nabla^d  \delta u^{ef}\;,\label{eq:deltauab}
\end{align}
where   the notation $\rho=\rho_0$ is adopted for simplicity and the thermodynamic quantities include the isothermal  expansion coefficient $\alpha \equiv -  \left({\partial\rho}/{\partial T}\right)_{p,y}/{\rho}$, the derivative of the pressure with respect to the vacancy molar fraction $\pi_y \equiv  \left({\partial p}/{\partial y}\right)_{\boldsymbol{\mathsf u},T}$, the isothermal bulk modulus $B_T \equiv \rho \left({\partial p}/{\partial \rho}\right)_{T,y}$,  the derivative of the specific entropy ${\mathfrak s}\equiv s/\rho$ (i.e., the entropy per unit mass)  with respect to the vacancy molar fraction $\varsigma_y \equiv \left({\partial{\mathfrak s}}/{\partial y}\right)_{\boldsymbol{\mathsf u},T}$, and the specific heat capacity at constant volume $c_v\equiv T\left({\partial{\mathfrak s}}/{\partial T}\right)_{\boldsymbol{\mathsf u},y}$. The elements of the tensor $(u^{ab})$ are denoted by  $\boldsymbol{\mathsf u}$, whereas $\boldsymbol{\mathsf u^\prime}$ denotes the elements other than the one involved in the partial derivative. The isothermal stress-strain coefficients $B_T^{abcd} \equiv \left({\partial \sigma^{ab}}/{\partial u^{cd}}\right)_{\boldsymbol{\mathsf u^\prime},T,y}$ and $G_T^{abcd} \equiv \left({\partial \phi^{ab}}/{\partial u^{cd}}\right)_{\boldsymbol{\mathsf u^\prime},T,y}$ characterize the elastic properties of the crystal. 

\subsection{Fourier-Laplace transform and decomposition in transverse and longitudinal sets of equations}
%%%%%%%%%%%%%%%%%%%%%%%%%%%%%%%%%%%%%%%%%%%%%%%%%%%%%%%%%%
\begin{table}[t!]
\begin{tabular}{c @{\hskip 1cm} c @{\hskip 1cm} c @{\hskip 1cm} c  }
\hline\hline
Direction&   $[100]$     &    $[110]$  &  $[111]$   	 \\
\hline  
${\bf e}_{\rm l}$ & ${\bf e}_x$  	& $({{\bf e}_x+{\bf e}_y})/{\sqrt{2}}$ 	& $({{\bf e}_x+{\bf e}_y+{\bf e}_z})/{\sqrt{3}}$	\\
$ {\bf e}_{{\rm t}_1}$		& $  {\bf e}_y$		& $({{\bf e}_x-{\bf e}_y})/{\sqrt{2}}$ & $ ({{\bf e}_x-{\bf e}_y})/{\sqrt{2}}$ \\
$ {\bf e}_{{\rm t}_2}$		& $  {\bf e}_z$		& $  {\bf e}_z$ & $({{\bf e}_x+{\bf e}_y-2{\bf e}_z})/{\sqrt{6}}$ \\
$B_{T,\rm l} $		& $B_{T,11}$		& $({B_{T,11}+B_{T,12}+2B_{T,44}})/{2}$ & $ ({B_{T,11}+2B_{T,12}+4B_{T,44}})/{3}$ \\
$ B_{T,{\rm t}_1}$		& $B_{T,44}$		& $({B_{T,11}-B_{T,12}})/{2}$ & $ ({B_{T,11}-B_{T,12}+B_{T,44}})/{3}$ \\
$ B_{T,{\rm t}_2}$		& $B_{T,44}$		& $B_{T,44}$ & $ ({B_{T,11}-B_{T,12}+B_{T,44}})/{3}$ \\
$G_{T,\rm l} $		& $2\left(B_{T,11}-B_{T,12}\right)/3$		& $(B_{T,11}-B_{T,12}+6B_{T,44})/6$ & $ 4B_{T,44}/{3}$ \\
$ G_{T,{\rm t}_1}$		& $B_{T,44}$		& $({B_{T,11}-B_{T,12}})/{2}$ & $ ({B_{T,11}-B_{T,12}+B_{T,44}})/{3}$ \\
$ G_{T,{\rm t}_2}$		& $B_{T,44}$		& $B_{T,44}$ & $ ({B_{T,11}-B_{T,12}+B_{T,44}})/{3}$ \\
$\eta_{\rm l} $		& $ \eta_{11}$		& $({\eta_{11}+\eta_{12}+2\eta_{44}})/{2}$ & $({\eta_{11}+2\eta_{12}+4\eta_{44}})/{3}$ \\
$ \eta_{{\rm t}_1}$		& $\eta_{44}$		& $({\eta_{11}-\eta_{12}})/{2}$ & $({\eta_{11}-\eta_{12}+\eta_{44}})/{3}$ \\
$ \eta_{{\rm t}_2}$		& $\eta_{44}$		& $\eta_{44}$ & $({\eta_{11}-\eta_{12}+\eta_{44}})/{3}$ \\                
\hline\hline
\end{tabular}
\caption{The stress-strain coefficients $B_{T,\sigma}$ and $G_{T,\sigma}$, and the viscosity coefficients $\eta_{\sigma}$ in the longitudinal and transverse directions ${\bf e}_\sigma$ with $\sigma\in\{{\rm l},{\rm t}_1,{\rm t}_2\}$ for the wave vector ${\bf q}$ oriented in the directions $[100]$, $[110]$, and $[111]$ of the crystalline lattice, as expressed using Voigt notations. The relations $B_T^{abcd}= B_T\delta^{ab}\delta^{cd}+G_T^{abcd}$, giving $G_{T,\sigma}=B_{T,\sigma}-B_T\delta_{{\rm l}\sigma}$, and  $B_T=(B_{T,11}+2B_{T,12})/3$~\cite{W98} are used to expressed  $G_{T,\sigma}$ in terms of the components of $B_T^{abcd}$. }
\label{Tab:LTCoeffs}
\end{table}
%%%%%%%%%%%%%%%%%%%%%%%%%%%%%%%%%%%%%%%%%%%%%%%%%%%%%%%%%%

We solve the closed system of equations in Fourier-Laplace space, where spatial and time derivatives are replaced by wave vectors ${\bf q}$ and frequencies $z$, respectively.  The Fourier-Laplace transform $\tilde{f}({\bf q},z)$ of the function $f({\bf r},t)$ is defined as
\begin{align}
\tilde{f}({\bf q},z)&\equiv\int_0^\infty \text{d}t\ {\rm e}^{-zt}\int_{V} \text{d}{\bf r}\ {\rm e}^{{\rm i}{\bf q}\cdot{ \bf r}}f({\bf r},t)=\int_0^\infty \text{d}t\ {\rm e}^{-zt}f({\bf q},t) \, ,\label{eq:FLT}
\end{align}
where $f({\bf q},t)$ is the Fourier transform of $f({\bf r},t)$. By expressing the closed set of equations~\eqref{eq:deltay}-\eqref{eq:deltauab} in this space (see Appendix~\ref{app:FLT} for details), we  select  preferred directions for the wave vector ${\bf q}$. For these directions, the system decomposes into independent sets of transverse and longitudinal equations, as shown in the following.

We define the orthonormal basis $\{{\bf e}_{\rm l},{\bf e}_{\rm t_1},{\bf e}_{\rm t_2} \}$, where the longitudinal direction is along the wave vector ${\bf q}=q{\bf e}_{\rm l}$, with $q\equiv\Vert{\bf q}\Vert$ is the wave number. The transverse directions $\rm t_1$ and $\rm t_2$ are orthogonal to ${\bf e}_{\rm l}$ such that ${\bf e}_{\rm l}\cdot{\bf e}_{\rm t_k}=0$ and ${\bf e}_{{\rm t}_k}\cdot{\bf e}_{{\rm t}_{k'}}=\delta_{kk'}$, where $k,k'=1,2$. In this basis, the strain-tensor becomes
\begin{align}
	u^{ab}({\bf q})=-\frac{{\rm i}}{2}qu_\sigma({\bf q})\left(e_{\rm l}^ae_\sigma^b+e_{\rm l}^be_\sigma^a\right)\,,\label{eq:uabq}
\end{align}
as shown in Eq.~\eqref{eq:uabqapp}, where $\sigma\in\{\rm l,t_1,t_2\}$, and the components $u_\sigma$ read
\begin{align}
	u_{\rm l}({\bf q})={\rm i}q^{-1}u^{ab}({\bf q})e^a_{\rm l}e^b_{\rm l}\,,&&u_{{\rm t}_k}({\bf q})&=2{\rm i}q^{-1}u^{ab}({\bf q})e^a_{\rm l}e^b_{{\rm t}_k}\;,
\end{align}
as shown in Eqs.~\eqref{eq:appulq} and~\eqref{eq:apputq}, respectively. Moreover, using $u^{aa}({\bf q})=-{\rm i}qu_{\rm l}({\bf q})$, as shown in Eq.~\eqref{eq:appuaaq}, and the relation~\eqref{eq:relrhoyuaa}, the longitudinal component  $\delta u_{\rm l}$ can be expressed in terms of $\delta \rho$ and $\delta y$ as 
\begin{align}
	-{\rm i}q\delta u_{\rm l}({\bf q})=-\delta y({\bf q})-\frac{\delta\rho({\bf q})}{\rho_{0}}\;.\label{eq:ul}
\end{align}
Similarly, the velocity is expressed as ${\bf v}=v_\sigma{\bf e}_\sigma $, with $v_\sigma = {\bf v}\cdot {\bf e}_\sigma$.

From the rank-four tensors for the viscosities and the stress-strain coefficients, we define the rank-two tensors:
\begin{align}
		   \eta_{\sigma\sigma'}\equiv\left(\eta^{abcd}e_{\rm l}^a e_{\rm l}^c\right)e_\sigma^be_{\sigma'}^d\;,&&B_{T,\sigma\sigma'}\equiv\left(B_T^{abcd}e_{\rm l}^a e_{\rm l}^c\right)e_\sigma^be_{\sigma'}^d\;, && G_{T,\sigma\sigma'}\equiv\left(G_T^{abcd}e_{\rm l}^a e_{\rm l}^c\right)e_\sigma^be_{\sigma'}^d\;.\label{eq:4tbasis}
\end{align}
For cubic symmetry, these tensors are diagonal in the special directions $[100], [110]$ and $[111]$ of the wave vector~\cite{MG24b}:
\begin{align}
	\eta_{\sigma\sigma'}=\eta_{\sigma}\delta_{\sigma\sigma'}\;,&&B_{T,\sigma\sigma'}=B_{T,\sigma}\delta_{\sigma\sigma'}\;,&&G_{T,\sigma\sigma'}=G_{T,\sigma}\delta_{\sigma\sigma'}\;,\label{eq:2tbasis}
\end{align}
(no summation over $\sigma$). The special directions and their corresponding coefficients are defined in Table~\ref{Tab:LTCoeffs}.

In the following, we derive the spectral functions for the wave vector along the special directions, where the set of equations decomposes into a first set of four longitudinal equations:
\begin{align}
	z\delta\tilde\rho({\bf q},z)-{\rm i}q\rho\delta \tilde  v_{\rm l}({\bf q},z)&=\delta\rho({\bf q},0)\;,\label{eq:lrho}\\
	-q^2\frac{G_{T,{\rm l}}}{\rho^2c_v}\left[ \xi+\zeta T \left(\rho\varsigma_y-\alpha B_T \right) \right]\delta\tilde\rho({\bf q},z)+\left\{z+\frac{q^2}{\rho c_v}\left[ \kappa^{\prime}+\xi^\prime\left(\rho\varsigma_y-\alpha B_T \right)\right] \right\}  \delta \tilde{T}({\bf q},z) &\notag\\
-{\rm i}q\frac{\alpha B_TT}{\rho c_v}\delta \tilde  v_{\rm l}({\bf q},z) -\frac{q^2}{\rho c_v}\left[ {G_{T,{\rm l}} \xi}+ {\cal K}_y+T\left( {\cal D}_y+\zeta G_{T,{\rm l}}\right) \left(\rho\varsigma_y-\alpha B_T \right)\right]  \delta\tilde {y}({\bf q},z)&=  \delta {T}({\bf q},0)\label{eq:lT}\,,\\
	-{\rm i}q\frac{B_{T,{\rm l}}}{\rho^2} \delta\tilde \rho({\bf q},z)-{\rm i}q  \frac{\alpha B_T }{\rho }  \delta \tilde{T}({\bf q},z)+\left(z+ q^2\frac{\eta_{\rm l}}{\rho}\right)\delta \tilde v_{\rm l}({\bf q},z)-{\rm i}q\frac{\pi_y+ B_{T,{\rm l}}}{\rho}\delta\tilde{y}({\bf q},z)&= \delta  v_{\rm l}({\bf q},0)\;,\label{eq:lv}\\
	q^2\frac{\zeta G_{T,{\rm l}}}{\rho}\delta\tilde\rho({\bf q},z)-q^2\frac{\xi'}{T}\delta \tilde{T}({\bf q},z)+\left[z+q^2\left( {\cal D}_y+\zeta G_{T,{\rm l}}\right)\right]\delta \tilde{y}({\bf q},z)&=\delta {y}({\bf q},0)\;,\label{eq:ly}
\end{align}
and two sets of two transverse equations
\begin{align}
	\left(z+ q^2\frac{\eta_{{\rm t}_k}}{\rho}\right)\delta \tilde v_{{\rm t}_k}({\bf q},z)+q^2\frac{B_{T,{\rm t}_k}}{\rho} \delta \tilde u_{{\rm t}_k}({\bf q},z)&=\delta  v_{{\rm t}_k}({\bf q},0)	\;,\label{eq:tv}\\
	-\delta \tilde{v}_{{\rm t}_k}({\bf q},z)+\left(z+q^2G_{T,{\rm t}_k} \zeta \right)  \delta \tilde{u}_{{\rm t}_k}({\bf q},z)&=\delta  u_{{\rm t}_k}({\bf q},0)\label{eq:tu}\;,
\end{align}
as shown in Appendix~\ref{app:TLE}.

In the absence of vacancies ($y=0$, $\zeta=0$, $\xi'=\xi=0$, $\kappa'=\kappa$), the seven equations describing perfect crystals are recovered.  In Ref.~\cite{MG24b} these simplified equations are used to derive the spectral functions for perfect crystals.

We solve the set of transverse equations by matrix inversion in Appendix~\ref{app:transverse} to obtain the spectral functions of transverse momentum fluctuations. The dispersion relations obtained from the poles of the spectral functions give two pairs of acoustic modes propagating with velocities
\begin{align} 
c_{{\rm t}_k\pm}\equiv\pm \sqrt{\frac{B_{T,{\rm t}_k}}{\rho}}\;,\label{eq:tcs}
\end{align}
 in the transverse directions ${\rm t}_k$ and with damping coefficients given by
\begin{align} 
\Gamma_{{\rm t}_k}\equiv\frac{1}{2}\left(\frac{\eta_{{\rm t}_k}}{\rho}+G_{T,{\rm t}_k} \zeta \right)\;,\label{eq:Gammatfull}
\end{align}
in agreement with the results of Refs.~\cite{FC76,MG25}. The expression for the propagation velocity of sound has the same form as for the perfect crystal. However, the velocity might be different if $B_{T,{\rm t}_k}$ changes in the presence of vacancies. The expression for the damping coefficient $\Gamma_{{\rm t}_k}$ differs from that of the perfect crystal by  a correction proportional to the molar fraction of vacancies, since  the vacancy conductivity $\zeta$ is linear in $y$~\cite{MG25}. To characterize the other four modes of the crystal, we solve the system of longitudinal equations.

%%%%%%%%%%%%%%%%%%%%%%%%%%%%%%%%%%%%%%%%%%%%%%%%%%%%%%%%%%%%
\section{The dynamic structure factor and its resonances} 
\label{sec:DSF}
%%%%%%%%%%%%%%%%%%%%%%%%%%%%%%%%%%%%%%%%%%%%%%%%%%%%%%%%%%%%
The dynamic structure factor $S({\bf q},\omega)$ is defined as the spectral function of the density fluctuations
\begin{align}
	S({\bf q},\omega)\equiv\frac{1}{Nm^2}\int_{\mathbb R} \text{d}t\ {\rm e}^{-{\rm i}\omega t}\left\langle\delta \hat \rho({\bf q},t)\delta \hat\rho^*({\bf q},0)\right\rangle_{\rm eq}\;,
\end{align} 
where $\delta \hat \rho$ is the microscopic mass density.  Using the property relating the Fourier and the Laplace transforms for real and even functions of time~\cite{BP76}, the dynamic structure factor is obtained from $\delta  \hat{\tilde{\rho}}({\bf q},z)$ by  the formula
\begin{align}
S({\bf q},\omega)  = \frac{2}{Nm^2}{\rm Re}\left[\langle\delta   \hat{\tilde{\rho}}({\bf q},z={\rm i}\omega)\delta \hat \rho^*({\bf q},0)\rangle_{\rm eq}\right].\label{eq:DSFFTFLT}
\end{align} 

In this section, we first derive the correlation function $\langle\delta   \hat{\tilde{\rho}}({\bf q},z)\delta \hat\rho^*({\bf q},0)\rangle_{\rm eq}$ by solving the longitudinal equations~\eqref{eq:lrho}-\eqref{eq:ly} and using Onsager's hypothesis of regression of fluctuations. From the resulting expression and Eq.~\eqref{eq:DSFFTFLT}, we obtain a formal and exact analytical expression for the dynamic structure factor. By finding the poles of the spectral function, we identify the dispersion relations of the modes and characterize the resonances of the DSF. Finally, with a small wave number expansion, we express the structure factor as a sum of Lorentzian functions.

\subsection{Full analytical expression for the spectrum}

The set of longitudinal equations~\eqref{eq:lrho}-\eqref{eq:ly} is cast in a matrix form as
\begin{align}\label{eq:M-phi=phi}
\boldsymbol{\mathsf M}_{\rm l}(q,z) \cdot \delta\tilde{\boldsymbol{\phi}}_{\rm l}({\bf q},z)= \delta\boldsymbol{\phi}_{\rm l}({\bf q},0) \, ,
\end{align}
where $\delta{\boldsymbol{\phi}}_{\rm l}\equiv(\delta\rho,\delta T,\delta   v_{\rm l}, \delta y)^{{\top}}$ and the matrix $\boldsymbol{\mathsf M}_{\rm l}$ reads
\begin{align}
\boldsymbol{\mathsf M}_{\rm l}(q,z) \equiv \left[
\begin{array}{cccc}
z &  0  & -{\rm i}\rho q &0  \\
-\frac{G_{T,{\rm l}}\left[ \xi+\zeta T \left(\rho\varsigma_y-\alpha B_T \right) \right]}{\rho^2c_v}q^2 & z+\frac{ \kappa^{\prime}+\xi^\prime\left(\rho\varsigma_y-\alpha B_T \right)}{\rho c_v}q^2 &-{\rm i}\frac{\gamma-1}{\alpha}q& -\frac{ {G_{T,{\rm l}} \xi}+ {\cal K}_y+T\left( {\cal D}_y+\zeta G_{T,{\rm l}}\right) \left(\rho\varsigma_y-\alpha B_T \right)}{\rho c_v}q^2\\
-{\rm i}\frac{B_{T,{\rm l}}}{\rho^2} q &-{\rm i}\frac{\alpha B_T }{\rho }    q& z+ \frac{\eta_{\rm l}}{\rho}q^2&-{\rm i}\frac{\pi_y+ B_{T,{\rm l}}}{\rho} q\\
\frac{\zeta G_{T,{\rm l}}}{\rho}q^2 & -\frac{\xi'}{T}q^2 & 0&z+\left( {\cal D}_y+\zeta G_{T,{\rm l}}\right)q^2 \\
\end{array}
\right]\;.\label{eq:lin-rho-T-vl-qz}
\end{align}
The system is solved by finding the  inverse matrix  $\boldsymbol{\mathsf M}^{-1}_{\rm l}(q,z)=\boldsymbol{\mathsf C}^{{\top}}_{M_{\rm l}}(q,z)/{\rm det}\ \boldsymbol{\mathsf M}_{\rm l}(q,z)$, where $\boldsymbol{\mathsf C}_{M_{\rm l}}(q,z)$ is the cofactor matrix. Using Onsager's hypothesis of regression of fluctuations~\cite{O31b}, we replace the deviations $\delta{\boldsymbol{\phi}}_{\rm l}$ of the fields with their microscopic expressions $\delta \hat{\boldsymbol{\phi}}_{\rm l}$ in order to compute correlation functions. Further details on the use of Onsager's hypothesis in this context are given in Ref.~\cite{MG23}. Since the variables $(\delta\rho,\delta T, \delta v_{\rm l})$ are statistically independent, the static correlation functions $\langle\delta \hat v_{\rm l}({\bf q},0)\delta \hat\rho^*({\bf q},0)\rangle_{\rm eq}$ and $\langle\delta \hat T({\bf q},0)\delta \hat\rho^*({\bf q},0)\rangle_{\rm eq}$ vanish. As a consequence, the spectral function of  density fluctuations is given in terms of only two components of the inverse matrix as
\begin{align}
\langle\delta \hat{\tilde{\rho}}({\bf q},z)\delta \hat\rho^*({\bf q},0)\rangle_{\rm eq}=\frac{1}{{\rm det}\ \boldsymbol{\mathsf M}_{\rm l}(q,z)}\left\{\left[\boldsymbol{\mathsf C}_{M_{\rm l}}(q,z)\right]_{1,1}\left\langle\delta \hat\rho({\bf q},0)\delta \hat\rho^*({\bf q},0)\right\rangle_{\rm eq}+\left[\boldsymbol{\mathsf C}_{M_{\rm l}}(q,z)\right]_{4,1}\left\langle\delta  \hat y({\bf q},0)\delta\hat \rho^*({\bf q},0)\right\rangle_{\rm eq}\right\}\;,\label{eq:Sks}
\end{align}
where the components $\left[\boldsymbol{\mathsf C}_{M_{\rm l}}(q,z)\right]_{1,1}$ and $\left[\boldsymbol{\mathsf C}_{M_{\rm l}}(q,z)\right]_{4,1}$ of the cofactor matrix, and the determinant ${\rm det}\ \boldsymbol{\mathsf M}_{\rm l}(q,z)$ are computed in Eqs.~\eqref{eq:C11},~\eqref{eq:C41} and~\eqref{eq:detMl}, respectively. In terms of the frequency $\omega = -{\rm i}z$, these quantities read
\begin{align}
	{\rm det}\ \boldsymbol{\mathsf M}_{\rm l}(q,z={\rm i}\omega)&= D_1(q,\omega)+{\rm i}D_2(q,\omega)\;,\label{eq:detMlomega}\\
	\left[\boldsymbol{\mathsf C}_{M_{\rm l}}(q,z={\rm i}\omega)\right]_{1,1}&= N_1(q,\omega)+{\rm i}N_2(q,\omega)\;,\\
	\left[\boldsymbol{\mathsf C}_{M_{\rm l}}(q,z={\rm i}\omega)\right]_{4,1}&= M_1(q,\omega)+{\rm i}M_2(q,\omega)\;,
\end{align}
with the coefficients
\begin{align}
	D_1(q,\omega)&\equiv\omega^4-\omega^2\left\{\frac{B_{T,{\rm l}} +B_T   \left({\gamma-1}\right)}{\rho}q^2\right.\notag\\
	&\left.+\left[\frac{ \kappa^{\prime}\eta_{\rm l} }{\rho^2 c_v}+\frac{ \kappa^{\prime} \left({\cal D}_y+\zeta G_{T,{\rm l}}\right)}{\rho c_v}+\frac{\eta_{\rm l}\left({\cal D}_y+\zeta G_{T,{\rm l}}\right)}{\rho}+\frac{\xi'\eta_{\rm l}\left(\rho\varsigma_y-\alpha B_T \right)}{\rho^2 c_v}-\frac{\xi'\left( {G_{T,{\rm l}} \xi}+ {\cal K}_y\right)}{T\rho c_v}\right]q^4\right\}\notag\\
	&+\frac{1}{\rho^2 c_v}\left[-{\cal K}_y {B_{T,{\rm l}}}\left(\frac{\xi'}{T}\right)+ {\cal D}_y{G_{T,{\rm l}} \xi}{\alpha B_T }- \kappa^{\prime} \pi_y \zeta G_{T,{\rm l}}+{\cal D}_y{ \kappa^{\prime}}{B_{T,{\rm l}}}-{\zeta G_{T,{\rm l}}}\alpha B_T  {\cal K}_y+{\xi}\pi_y G_{T,{\rm l}}\left(\frac{\xi'}{T}\right) \right]q^6\;,\label{eq:D1}\\
	D_2(q,\omega)&\equiv \omega\left\{-\omega^2 \left[\frac{ \kappa^{\prime} }{\rho c_v}+ \frac{\eta_{\rm l}}{\rho}+{\cal D}_y+\zeta G_{T,{\rm l}}+\frac{\xi^\prime\left(\rho\varsigma_y-\alpha B_T \right) }{\rho c_v}\right]q^2\right.\notag\\
	&+\left[{\cal D}_y\frac{{B_{T,{\rm l}}} +B_T   \left({\gamma-1}\right)}{\rho}+ \frac{ \kappa^{\prime}B_{T,{\rm l}}}{\rho^2 c_v}+\frac{\pi_y}{\rho}\left(\frac{\alpha B_T\xi'}{\rho c_v }- \zeta G_{T,{\rm l}} \right)+\frac{{G_{T,{\rm l}} \xi}{\alpha B_T }}{\rho^2c_v}+\varsigma_y\frac{\xi' B_{T,{\rm l}}+\zeta T \alpha B_T G_{T,{\rm l}}}{\rho c_v}\right]q^4\notag\\
	&\left.+\left[ \frac{ \kappa^{\prime}\eta_{\rm l}\left({\cal D}_y+\zeta G_{T,{\rm l}}\right) }{\rho^2 c_v}-\frac{\eta_{\rm l}\xi'\left( {G_{T,{\rm l}} \xi}+ {\cal K}_y\right)}{T\rho^2 c_v}\right]q^6\right\}\;,\label{eq:D2}\\
	N_1(q,\omega)&\equiv-\omega^2\left[\frac{ \kappa^{\prime} }{\rho c_v}+ \frac{\eta_{\rm l}}{\rho}+{\cal D}_y+\zeta G_{T,{\rm l}}+\frac{\xi^\prime\left(\rho\varsigma_y-\alpha B_T \right) }{\rho c_v}\right]q^2\notag\\
	&+\left[\frac{\left(\gamma-1\right)\left(\pi_y+ B_{T,{\rm l}}\right)\xi'}{\alpha\rho T }+\frac{ \left( {\cal D}_y+\zeta G_{T,{\rm l}}\right)B_T   \left({\gamma-1}\right) }{\rho }\right]q^4\notag\\
	&+\left[\frac{ \kappa^{\prime}\eta_{\rm l}\left({\cal D}_y+\zeta G_{T,{\rm l}}\right) }{\rho^2 c_v}-\frac{\eta_{\rm l}\xi'\left( {G_{T,{\rm l}} \xi}+ {\cal K}_y\right)}{T\rho^2 c_v}\right]q^6\;,\\
	N_2(q,\omega)&\equiv\omega\left\{-\omega^2+\frac{B_T   \left({\gamma-1}\right)}{\rho}q^2\right.\notag\\
	&\left.+\left[\frac{ \kappa^{\prime}\eta_{\rm l} }{\rho^2 c_v}+\frac{ \kappa^{\prime} \left({\cal D}_y+\zeta G_{T,{\rm l}}\right)}{\rho c_v}+\frac{\eta_{\rm l}\left({\cal D}_y+\zeta G_{T,{\rm l}}\right)}{\rho}+\frac{\xi'\eta_{\rm l}\left(\rho\varsigma_y-\alpha B_T \right)}{\rho^2 c_v}-\frac{\xi'\left( {G_{T,{\rm l}} \xi}+ {\cal K}_y\right)}{T\rho c_v}\right]q^4\right\}\;,\\
	M_1(q,\omega)&\equiv -\frac{1}{\rho c_v}\left\{{\alpha B_T}\left[{G_{T,{\rm l}} \xi}+ {\cal K}_y+T\left( {\cal D}_y+\zeta G_{T,{\rm l}}\right) \left(\rho\varsigma_y-\alpha B_T \right)\right] +\left[\kappa^{\prime}+\xi^\prime\left(\rho\varsigma_y-\alpha B_T \right) \right]\left({\pi_y+ B_{T,{\rm l}}}\right) \right\}q^4\;,\\
	M_2(q,\omega)&\equiv-\omega \left({\pi_y+ B_{T,{\rm l}}}\right)q^2\label{eq:M2}\;.
\end{align}

Substituting Eq.~\eqref{eq:Sks} into Eq.~\eqref{eq:DSFFTFLT}, the dynamic structure factor is finally found as 
\begin{align}
	S({\bf q},\omega) &=2S({\bf q}) \frac{N_1(q,\omega)D_1(q,\omega)+N_2(q,\omega)D_2(q,\omega)}{D^2_1(q,\omega)+D^2_2(q,\omega)}+2{S_{\rm nv}({\bf q})} \frac{M_1(q,\omega)D_1(q,\omega)+M_2(q,\omega)D_2(q,\omega)}{D^2_1(q,\omega)+D^2_2(q,\omega)}\;,\label{eq:DSF_Full}
\end{align}
where $S({\bf q})\equiv\frac{1}{Nm^2}\langle\delta \hat\rho({\bf q},0)\delta\hat \rho^*({\bf q},0)\rangle_{\rm eq} $  is the static structure factor, which corresponds to the equal-time correlation function of density fluctuations. Similarly, $S_{\rm nv}({\bf q})\equiv\frac{1}{Nm^2}\langle\delta \hat y({\bf q},0)\delta \hat \rho^*({\bf q},0)\rangle_{\rm eq} $ is the static correlation function between the vacancy molar fraction and the particle density fluctuations. 
%%%%%%%%%%%%%%%%%%%%%%%%%%%%%%%%%%%%%%%%%%%%%%%%%%%%%%%%%%%%%%%%%%
\begin{figure}[t!]\centering
{\includegraphics[width=.75\textwidth]{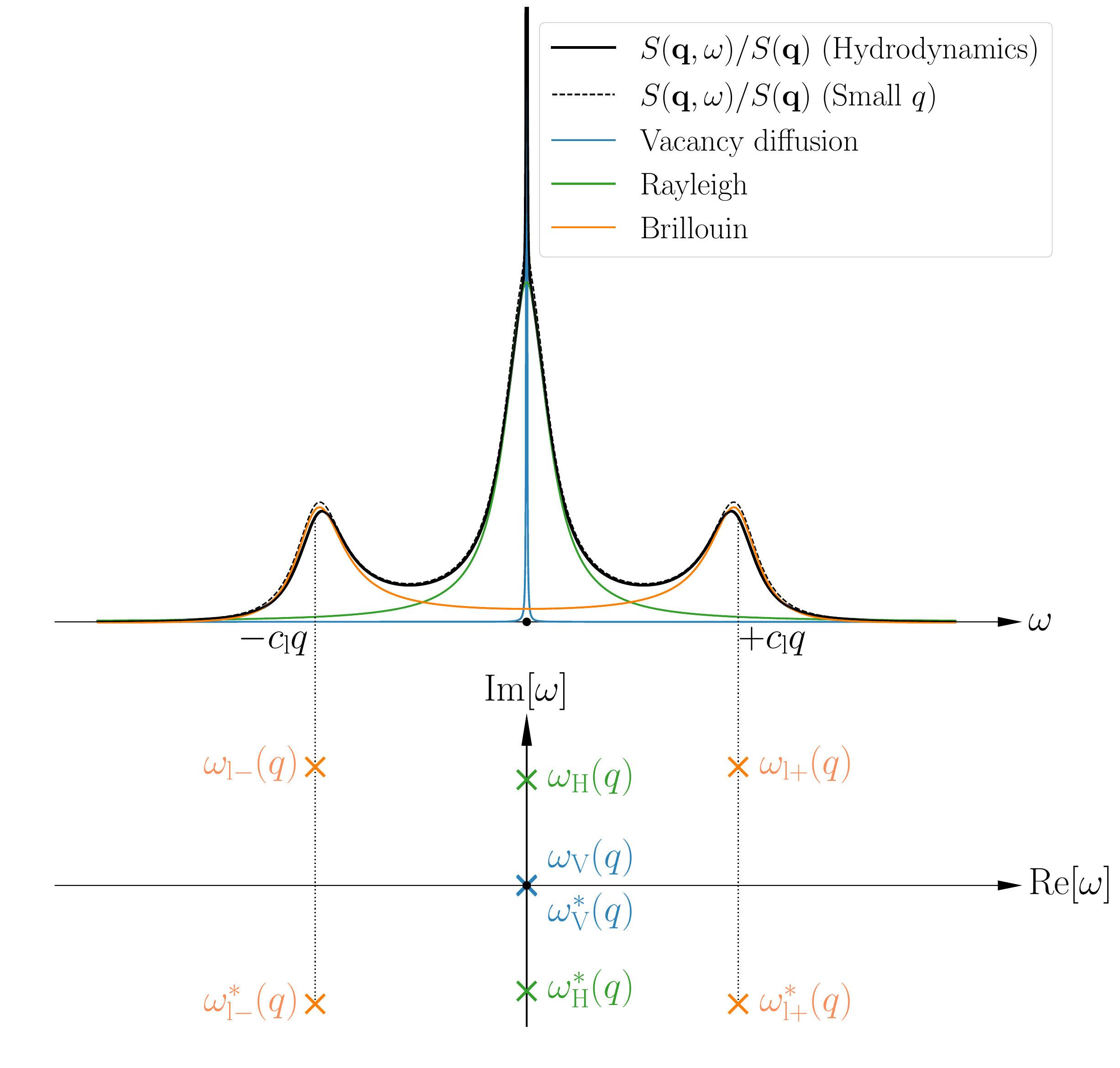}}
\caption[] {Representation of the dynamic structure factor $S({\bf q},\omega)$ for a cubic monatomic crystal with vacancies, at a wave number value of $q_*=\Vert{\bf q}_*\Vert=0.4$. The solid black line represents the analytical expression, Eq.~\eqref{eq:DSF_Full}, and the dashed black line shows the small $q$ approximation, Eq.~\eqref{eq:DSF_approx}. The colored solid lines represent the peaks identified in Eq.~\eqref{eq:DSF_approx}: vacancy diffusion (blue), heat diffusion (Rayleigh -- green), and sound propagation and damping (Brillouin -- orange). Below, their underlying poles are depicted in the plane of complex frequencies. The parameters are for a hard-sphere crystal of $499$ particles on $500$ lattice sites at a density $n_{0*}=1.05$ (in dimensionless units of the simulation), as given in Ref.~\cite{MG25}, and the ratio $S_{\rm nv}({\bf q})/S({\bf q})$ is taken as $-30/N_0$. For illustrative purposes, the central sharp peak is truncated vertically. } \label{Fig:DSF_Theo}
\end{figure}
%%%%%%%%%%%%%%%%%%%%%%%%%%%%%%%%%%%%%%%%%%%%%%%%%%%%%%%%%%%%%%%%%%

Equation~\eqref{eq:DSF_Full} is an exact analytical expression for the dynamic structure factor of a cubic monatomic crystalline solid with vacancies. This is the central result of this work.  The expression is given in terms of the thermodynamic, elastic, and transport properties of the system and stems from the transport equations~\eqref{eq:drho}-\eqref{eq:du}. This formula generalizes the result for a perfect cubic crystal found previously in Ref.~\cite{MG24b} to the case of a crystal with vacancies. In the absence of vacancies, both formulas are identical.

The dynamic structure factor is  depicted in Fig.~\ref{Fig:DSF_Theo}. It shows a structure similar to the perfect crystal, with one broad resonance at zero frequency (a purely diffusive mode) and a pair of resonances at finite frequencies (propagating modes). In contrast to a perfect crystal, in the presence of vacancies, the DSF has an additional, sharp resonance at zero frequency, which corresponds to a second diffusive mode. The identification and interpretation of these resonances are done in the next section.

\subsection{Resonances}
To identify and characterize the resonances of the dynamic structure factor from its analytical expression, Eq.~\eqref{eq:DSF_Full}, we first compute its poles, which are given by the roots of the determinant of the matrix $\boldsymbol{\mathsf M}_{\rm l}(q,z)$. This allows us to find the dispersion relations associated with the slow modes and associate them with the resonances  of the DSF. 

From the roots of the determinant, computed in Appendix~\ref{app:poles_DSF}, we identify four dispersion relations
\begin{align}
	z_{\rm H}&=-\chi q^2+\cdots&&\text{heat diffusion mode,}\label{eq:z_H}\\
	z_{\rm V}&=-D_{\rm vac}q^2+\cdots&&\text{vacancy diffusion mode,}\label{eq:z_V}\\
	z_{\rm l,\pm}&=\pm{\rm i} c_{\rm l}q-\Gamma_{\rm l} q^2+\cdots&&\text{longitudinal sound modes,}\label{eq:z_pm}
\end{align}
where the dots denote terms vanishing faster than $q^2$ in the limit $q\to 0$. The velocity of the longitudinal acoustic modes is
\begin{align}
c_{\rm l}\equiv\sqrt{\left[{B_{T,{\rm l}}} +B_T   \left({\gamma-1}\right)\right]/\rho}\;,\label{eq:lcs}
\end{align} 
and 
\begin{align}
	\Gamma_{\rm l}\equiv\frac{1}{2}\left[\frac{\eta_{\rm l}}{\rho}+\frac{ \kappa^{\prime} }{\rho c_v}\left(1-\frac{ B_{T,{\rm l}}}{\rho c_{\rm l}^2}\right)-\frac{ G_{T,{\rm l}} \xi \alpha B_T }{\rho^2c_vc_{\rm l}^2}+\left(\zeta G_{T,{\rm l}}-\frac{\xi^\prime\alpha B_T}{\rho c_v}\right)\left(1+\frac{\pi_y}{\rho c_{\rm l}^2}-\frac{\varsigma_yT\alpha B_T}{\rho c_v c_{\rm l}^2}\right)\right]
\label{eq:Gammalfull}
\end{align}
is the longitudinal sound damping coefficient.  The coefficients
\begin{align}
	\chi&\equiv\frac{C_{1,4}+\sqrt{C^2_{1,4}-4c_{\rm l}^2C_{0,6}}}{2c_{\rm l}^2}\;,\label{eq:chifull}\\
	D_{\rm vac}&\equiv \frac{C_{1,4}-\sqrt{C^2_{1,4}-4c_{\rm l}^2C_{0,6}}}{2c_{\rm l}^2}\;,\label{eq:Dvacfull}
\end{align}
are the heat and vacancy diffusivities, respectively, where
\begin{align}
	C_{1,4}&\equiv   {\cal D}_y\frac{{B_{T,{\rm l}}} +B_T   \left({\gamma-1}\right)}{\rho}+ \frac{ \kappa^{\prime}B_{T,{\rm l}}}{\rho^2 c_v}+\frac{\pi_y}{\rho}\left(\frac{\alpha B_T\xi'}{\rho c_v }- \zeta G_{T,{\rm l}} \right)+\frac{{G_{T,{\rm l}} \xi}{\alpha B_T }}{\rho^2c_v}+\varsigma_y\frac{\xi' B_{T,{\rm l}}+\zeta T \alpha B_T G_{T,{\rm l}}}{\rho c_v} \;,\\
		C_{0,6}&\equiv\frac{1}{\rho^2 c_v}\left[{\cal D}_y{ \kappa^{\prime}}{B_{T,{\rm l}}}-{\cal K}_y {B_{T,{\rm l}}}\left(\frac{\xi'}{T}\right)+ {\cal D}_y{G_{T,{\rm l}} \xi}{\alpha B_T }- \kappa^{\prime} \pi_y \zeta G_{T,{\rm l}}-{\zeta G_{T,{\rm l}}}\alpha B_T  {\cal K}_y+{\xi}\pi_y G_{T,{\rm l}}\left(\frac{\xi'}{T}\right) \right]\;,
\end{align} 
[with a notation justified by Eq.~\eqref{eq:detMl}]. The analytical expressions~\eqref{eq:lcs},~\eqref{eq:Gammalfull},~\eqref{eq:chifull}, and~\eqref{eq:Dvacfull}, for the longitudinal speed and damping coefficient of sound, and the heat and vacancy diffusivities, respectively, which include the contributions due to the presence of vacancies,  are the second central result of this work. 

The poles are shown at the bottom of Fig.~\ref{Fig:DSF_Theo}. The poles $\omega_{\rm l, \pm}$ (and their complex conjugates), which have a real part at  $\pm c_{\rm l}q$, correspond to a pair of longitudinal sound waves. These appear in the DSF as a Brillouin doublet at finite frequencies. The pole $\omega_{\rm H}$ (and its complex conjugate) at zero frequency has a larger imaginary part than the pole $\omega_{\rm V}$ (and its complex conjugate), so it corresponds to the central broad resonance of the DSF and is identified as a Rayleigh peak of heat diffusion. Finally, the pole $\omega_{\rm V}$ (and its complex conjugate) corresponds to the central sharp resonance and is related to the diffusion of vacancies. In this way, all the resonances of the DSF are related to dispersion relations and transport processes.

Together with the four modes from the transverse equations, we have identified the eight predicted modes of the crystalline solid. Compared to a perfect crystal, there is an eighth dispersion relation, $z_{\rm V}$. This mode is purely diffusive, with a diffusivity given by $D_{\rm vac}={\cal D}_y+{\cal O}(y)$, which is associated with the diffusion of vacancies.  This result confirms the microscopic theory of vacancy diffusion in crystals developed in Ref.~\cite{MG25}, which shows that the eighth mode is indeed a purely diffusive mode of transport of vacancies and has a diffusion coefficient given by ${\cal D}_y$.

The other dispersion relations are the same as for a perfect crystal when the molar fraction of vacancies vanishes. In this limit, the heat diffusivity and the longitudinal sound damping coefficient are given by $\chi_{0}\equiv\frac{ \kappa B_{T,{\rm l}}}{c_{\rm l}^2\rho^2 c_v}$ and $\Gamma_{\rm l 0}\equiv \frac{1}{2}\left[ \frac{\eta_{\rm l}}{\rho}+\frac{ \kappa }{\rho c_p}\frac{\gamma}{1+  \left({\gamma-1}\right)^{-1}{B_{T,{\rm l}}}/{B_T}}\right]$, respectively, in agreement with their perfect crystal expressions found in Ref.~\cite{MG24b}.  In the regime  $1\ll N_v \ll N \sim N_0$, the molar fraction of vacancies satisfies $y\ll1$. In this case, the corrections to the resonances of the perfect crystal due to the presence of vacancies are subdominant. In this regime, the dynamic structure factor of the crystal with vacancies is approximately given by the DSF of the perfect crystal, plus an additional sharp resonance at $\omega=0$ due to the diffusion of vacancies. This analysis should be confirmed by expanding the analytic formula in the limit of small $q$. Finally, note that the order ${\cal O}(y)$ corrections are obtained from the full expressions~\eqref{eq:Gammalfull},~\eqref{eq:chifull}, and~\eqref{eq:Dvacfull}.

\subsection{Approximate expression for the spectrum}
To characterize the  resonances of the dynamic structure factor, we take the small $q$ expansion of the inverse Laplace transform of Eq.~\eqref{eq:Sks}, followed by a Fourier transform to frequency space. 

The determinant of the matrix $\boldsymbol{\mathsf M}_{\rm l}$ can be written as
\begin{align}
	\det \boldsymbol{\mathsf M}_{\rm l}(q,z)=(z-z_{\rm H})(z-z_{\rm V})(z-z_{\rm l+})(z-z_{\rm l-})\;,
\end{align}
where $z_{\rm H}$, $z_{\rm V}$ and $z_{\rm l\pm}$ are the roots, given by Eqs.~\eqref{eq:z_H}-\eqref{eq:z_pm}. The derivation, detailed in Appendix~\ref{app:resonances}, gives the intermediate scattering function $F({\bf q},t)\equiv\frac{1}{Nm^2}\langle   \delta\hat\rho({\bf q},t)    \delta\hat\rho^*({\bf q},0)  \rangle_{\rm eq} $  as
\begin{align}
	\frac{F({\bf q},t)}{S({\bf q})}	&=\frac{1}{1+(\gamma-1)\frac{B_T}{B_{T,{\rm l}}}}\left\{\frac{\gamma-1}{B_{T,{\rm l}}}\left[D_{\rm l}(y)+ \mathcal{O}(y^2)+\mathcal{O}(q^2)\right]{\rm e}^{-D_{\rm vac}q^2|t|}+\frac{\gamma-1}{B_{T,{\rm l}}}\left[B_T+\mathcal{O}(y)+\mathcal{O}(q^2)\right]{\rm e}^{-\chi q^2|t|}\right.\notag\\
	&+\left.\left[\cos(qc_{\rm l}|t|)+\frac{3\Gamma_{\rm l 0}-D_v}{c_{\rm l}}q\sin(qc_{\rm l}|t|)+\mathcal{O}(y)+\mathcal{O}(q^2)\right]{\rm e}^{-\Gamma_{\rm l } q^2|t|}\right\}\;,\label{eq:ISF}
\end{align}
where $D_v\equiv\eta_{\rm l}/\rho$ is the longitudinal kinematic viscosity, and the coefficient multiplying the vacancy diffusion  mode is given by
\begin{align}
	D_{\rm l}(y)&\equiv\frac{\left(\pi_y+ B_{T,{\rm l}}\right)\xi}{\alpha T (\chi_{0}-{\cal D}_y)}+\frac{( \zeta G_{T,{\rm l}}-\Delta D_{\rm vac})B_T}{\chi_{0}-{\cal D}_y}-\frac{\rho}{\alpha T}\left[\frac{{\cal K}_y+T {\cal D}_y\left(\rho\varsigma_y-\alpha B_T \right)}{\chi_{0}-{\cal D}_y}+\frac{ \left({\kappa}-{\rho c_v }{\cal D}_y \right)\left({\pi_y+ B_{T,{\rm l}}}\right)}{\alpha B_T (\chi_{0}-{\cal D}_y)}\right]\frac{S_{\rm nv}({\bf q})}{S({\bf q})}\;,\label{eq:coeffDofy}
\end{align}
where $\Delta D_{\rm vac} \equiv D_{\rm vac}-{\cal D}_y$.

\begin{figure}[t!] \centering
    \begin{minipage}{0.48\textwidth}
        \centering
        \includegraphics[width=1.\textwidth]{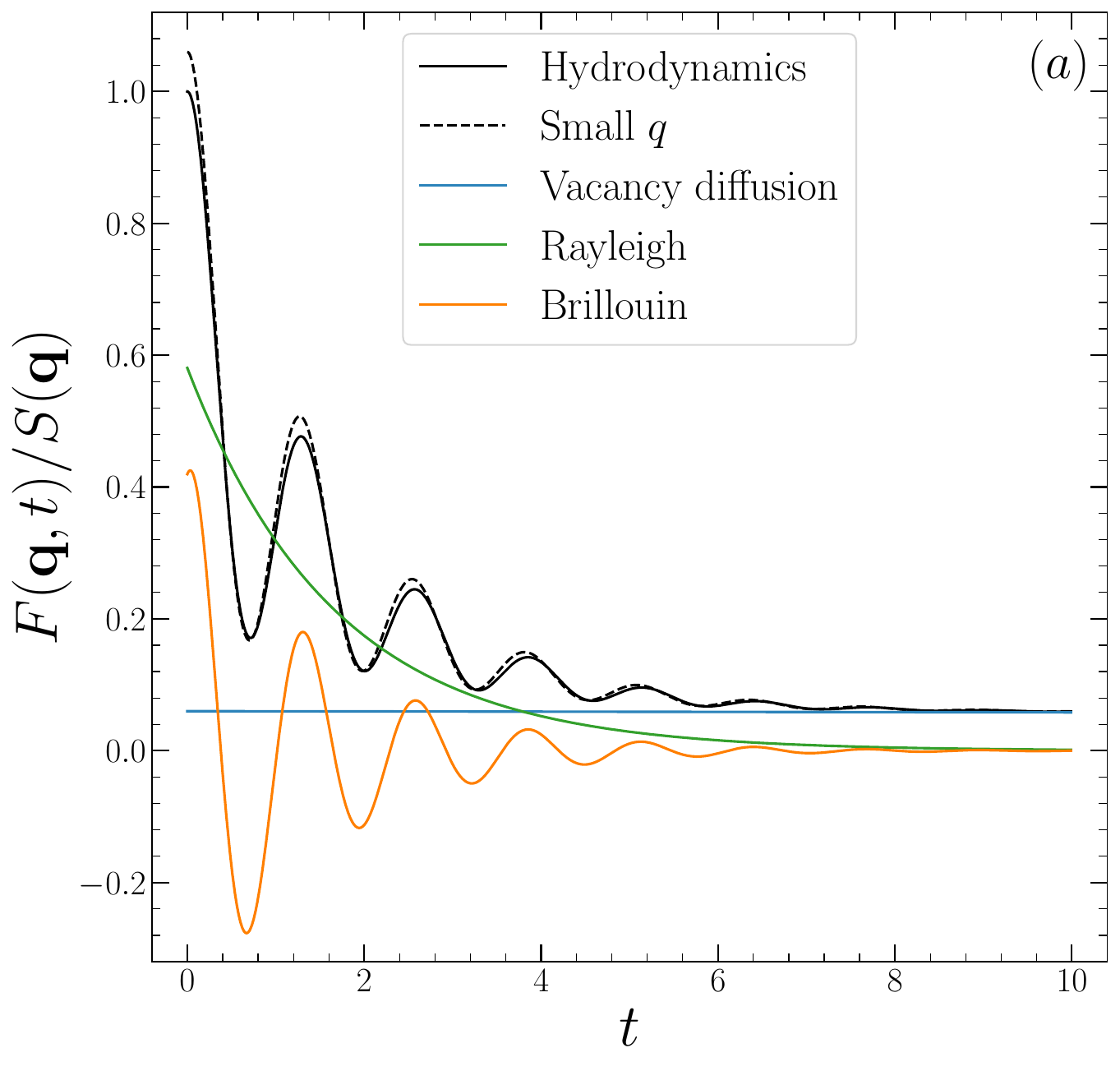} % second figure itself
    \end{minipage}\hfill
     \begin{minipage}{0.48\textwidth}
        \centering
        \includegraphics[width=1.\textwidth]{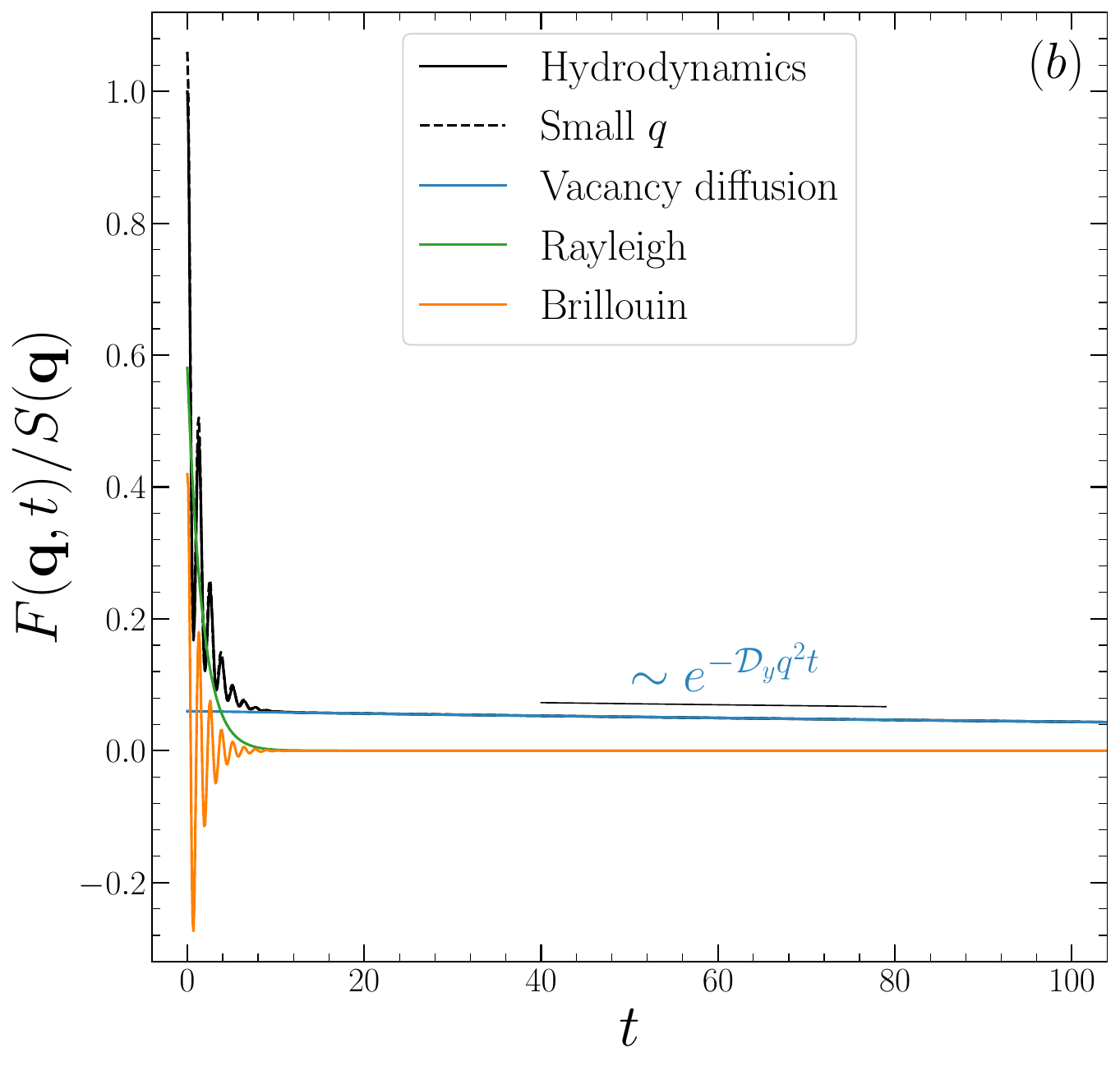} % first figure itself
    \end{minipage}
\caption[] {Representations of the intermediate scattering function, $F({\bf q},t)$, for a cubic monatomic crystal with vacancies at short times [panel~(a), left] and long times [panel~(b), right], for a wave vector value of  $q_*=\Vert{\bf q}_*\Vert=0.4$. The solid black line shows the inverse Fourier transform from frequency $\omega$ to time $t$ of the analytical expression, Eq.~\eqref{eq:DSF_Full}. The dashed black line represents the small $q$  approximation, Eq.~\eqref{eq:ISF}. The colored solid lines are the damping functions identified in Eq.~\eqref{eq:ISF}: vacancy diffusion (blue), heat diffusion (green), and sound propagation and damping (orange). The same parameters as in Fig.~\ref{Fig:DSF_Theo} are used. }\label{Fig:ISF_Theo}
\end{figure}

The intermediate scattering function~\eqref{eq:ISF},  depicted in Fig.~\ref{Fig:ISF_Theo}, is composed of four decaying modes. Two of these modes are purely diffusive, with damping factors that depend on $\chi$ and ${D}_{\rm vac}$, respectively. They  correspond to the heat and vacancy diffusion modes, with dispersion relations given by Eqs.~\eqref{eq:z_H} and~\eqref{eq:z_V}.  The other two modes are propagating, as indicated by  the oscillating functions with the longitudinal speed of sound $c_{\rm l}$ in their argument.  These modes  have a damping factor that depends on the acoustic attenuation coefficient $\Gamma_{\rm l}$. They are therefore the two longitudinal acoustic modes, with the dispersion relation given by Eq.~\eqref{eq:z_pm}.

The coefficient $D_{\rm l}$, defined in Eq.~\eqref{eq:coeffDofy}, which multiplies the vacancy diffusion mode,  is linear in the molar fraction of vacancies at leading order. This is  because $\zeta$, $\xi$, $\Delta D_{\rm vac}$, and $S_{\rm nv}({\bf q})$ are of order ${\cal O}(y)$ and the other coefficients are of order ${\cal O}(1)$ in the limit $y\to 0$.  The facts that $\lim_{{\bf q}\to 0} S_{\rm nv}({\bf q})=\mathcal{O}(y)$ and $\lim_{{\bf q}\to 0} S({\bf q})=\mathcal{O}(1)$ are proved in Appendix~\ref{app:static-fns}.  On short time scales, the function $F({\bf q},t)$ is dominated by the heat and acoustic modes, and has a very similar shape to that of a perfect crystal. Since the vacancy diffusivity ${D}_{\rm vac}$, is typically much smaller than the  diffusivity of the heat mode $\chi$, and the sound damping coefficient $\Gamma_{\rm l}$, as shown in Ref.~\cite{MG25}, the mode associated with the diffusion of vacancies is the slowest to decay. At intermediate-time scales, once the heat and acoustic modes have decayed, the intermediate scattering function is dominated by the contribution from vacancy diffusion and its prefactor, $D_{\rm l}$, as shown in panel~(b) of Fig.~\ref{Fig:ISF_Theo}. Therefore, by estimating the  decay of the function in this time regime, it is possible to obtain the vacancy diffusion coefficient, ${\cal{D}}_y$, solely from the correlation of density fluctuations, without locating the actual positions of the vacancies, as explained in Sec.~\ref{sec:MDHS}. 

In the hydrodynamic regime, the intermediate scattering function~\eqref{eq:ISF} agrees with the inverse Fourier transform of the exact dynamical structure factor in Eq.~\eqref{eq:DSF_Full}, as seen in Fig.~\ref{Fig:ISF_Theo}. This agreement supports the formula~\eqref{eq:coeffDofy} for the coefficient $D_{\rm l}$. Since this coefficient is of order  ${\cal O}(y)$, the intermediate scattering function of a perfect crystal is recovered when the molar fraction of vacancies vanishes. We also observe in panel~(a) of Fig.~\ref{Fig:ISF_Theo} that the approximate expression~\eqref{eq:ISF} slightly differs from the inverse Fourier transform of the exact expression~\eqref{eq:DSF_Full}  at very small times. The latter correctly gives $F({\bf q},t=0)=S({\bf q})$. This behavior illustrates that for the approximate expression, the $\mathcal{O}(y)$ corrections to the heat and acoustic modes become relevant  in the limit $t\rightarrow 0$,  exactly canceling  the contribution from vacancy diffusion to recover the static structure factor $S({\bf q})$ at $t=0$.

%%%%%%%%%%%%%%%%%%%%%%%%%%%%%%%%%%%%%%%%%%%%%%%%%%%%%%%%%%%%%%%%
%%%%%%%%%%%%%%%%%%%%%%%%%%%%%%%%%%%%%%%%%%%%%%%%%%%%%%%%%%%%%%%%

The Fourier transform of the intermediate scattering function~\eqref{eq:ISF} from time to frequency gives the approximate expression for the dynamic structure factor
\begin{align}
	\frac{S({\bf q},\omega)}{S({\bf q})} & = \frac{1}{1+(\gamma-1)\frac{B_T}{B_{T,{\rm l}}}}\left\{\frac{\gamma-1}{B_{T,{\rm l}}}D_{\rm l}(y)\frac{2D_{\rm vac}q^2}{\omega^2+(D_{\rm vac}q^2)^2}
	+\frac{\gamma-1}{B_{T,{\rm l}}}B_T\frac{2\chi q^2}{\omega^2+(\chi q^2)^2}\right.\notag\\
	&\left.+\frac{\Gamma_{\rm l } q^2}{(\omega-c_{\rm l} q)^2+(\Gamma_{\rm l } q^2)^2}+\frac{\Gamma_{\rm l } q^2}{(\omega+c_{\rm l} q)^2+(\Gamma_{\rm l } q^2)^2}+\frac{3\Gamma_{\rm l 0}-D_v}{c_{\rm l}}q\left[\frac{\omega+c_{\rm l}q}{(\omega+c_{\rm l} q)^2+(\Gamma_{\rm l } q^2)^2}-\frac{\omega-c_{\rm l}q}{(\omega-c_{\rm l} q)^2+(\Gamma_{\rm l } q^2)^2}\right]\right\}\;.\label{eq:DSF_approx}
\end{align}
The result is depicted in Fig.~\ref{Fig:DSF_Theo} (dashed black line) and is compared to the exact expression given in Eq.~\eqref{eq:DSF_Full}, showing  excellent agreement for  a wave number $q$ in the hydrodynamic limit. The four decaying modes identified in the intermediate scattering function~\eqref{eq:ISF} become Lorentzian functions in the DSF.  Using the same arguments as for $F({\bf q},t)$, the dynamic structure factor of a perfect crystal is found  in the limit where the molar fraction of vacancies vanishes.   The derivation of the approximate expression for the DSF, Eq.~\eqref{eq:DSF_approx}, together with the coefficient $D_{\rm l}$, Eq.~\eqref{eq:coeffDofy}, is the third main result of this work.

\begin{figure}[t!] \centering
    \begin{minipage}{0.48\textwidth}
        \centering
        \includegraphics[width=1.\textwidth]{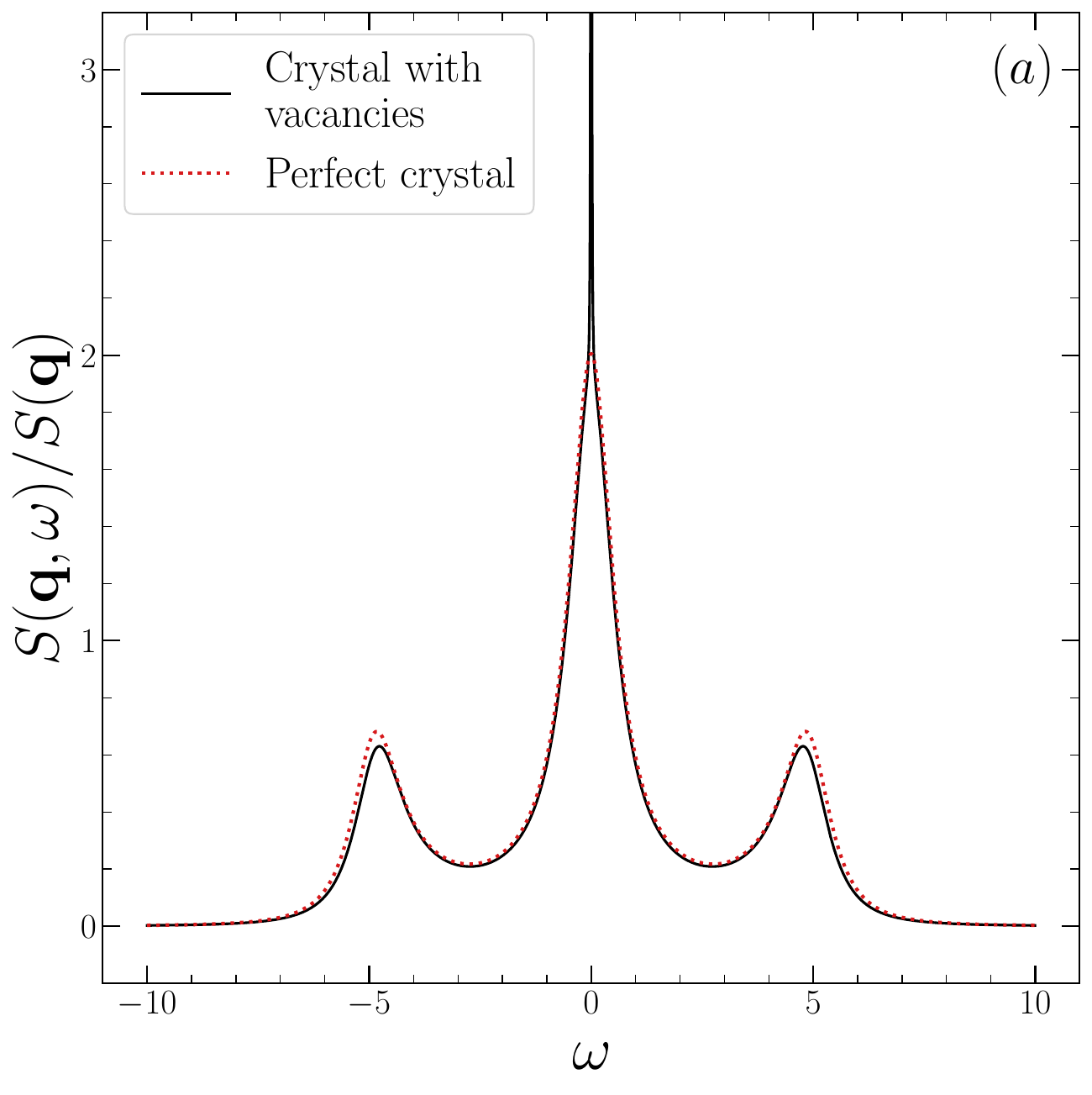} % second figure itself
    \end{minipage}\hfill
     \begin{minipage}{0.48\textwidth}
        \centering
        \includegraphics[width=1.\textwidth]{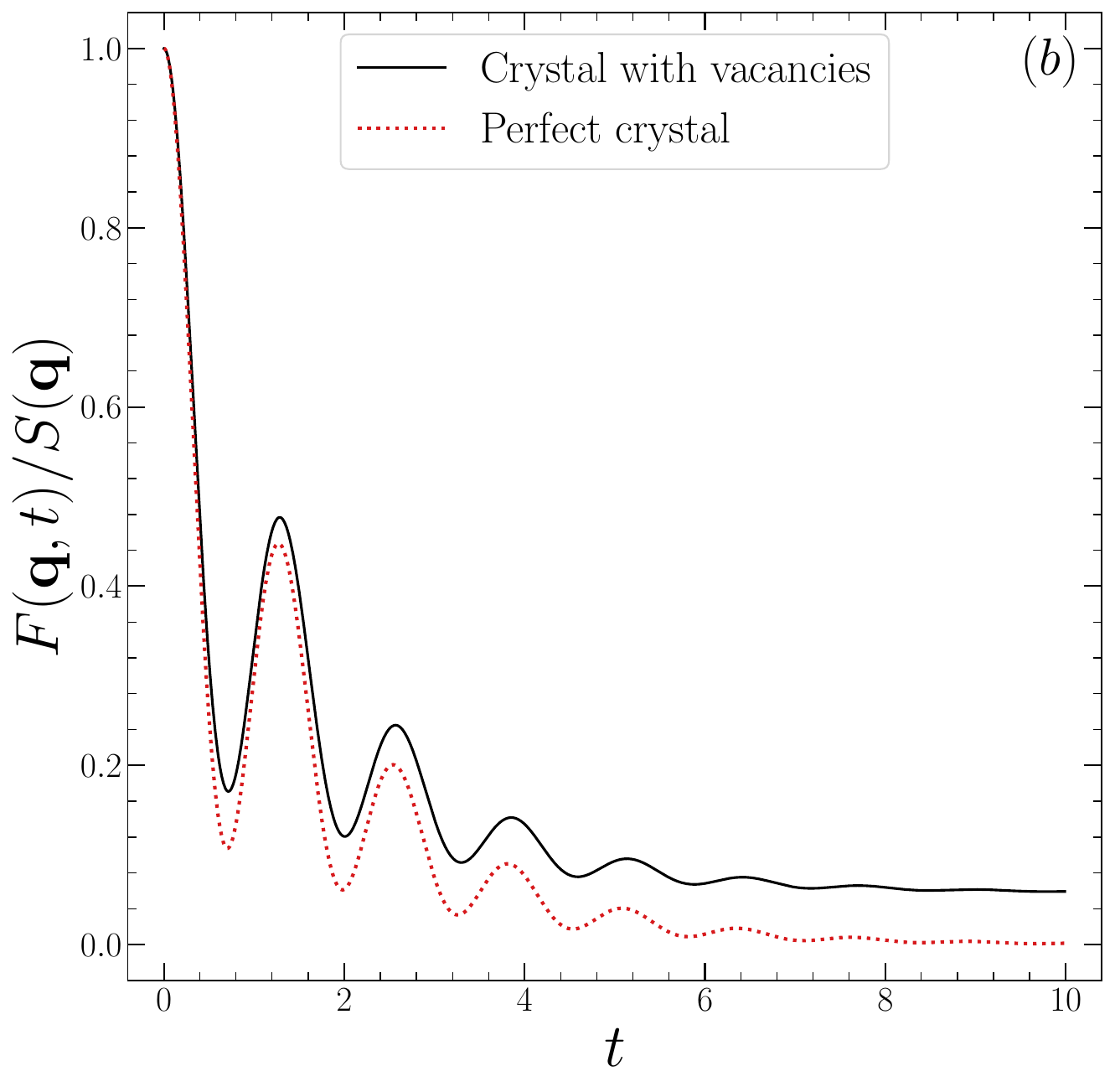} % first figure itself
    \end{minipage}
\caption[] {Representations of the dynamic structure factor, $S({\bf q},\omega)$, [panel~(a), left] and the intermediate scattering function, $F({\bf q},t)$, [panel~(b), right] for a cubic monatomic crystal with one vacancy (solid black lines) and without vacancies (perfect crystal -- dotted red lines) for a given value of the wave vector $q_*=\Vert{\bf q}_*\Vert=0.4$. The same parameters as in Fig.~\ref{Fig:DSF_Theo} are used. For illustrative purposes, the central sharp peak is truncated vertically in panel~(a).  }\label{Fig:perfect_vs_vac}
\end{figure}

 In Fig.~\ref{Fig:perfect_vs_vac}, the dynamic structure factors and  intermediate scattering functions of a crystal with a finite molar fraction of vacancies and a perfect crystal are compared in the hydrodynamic limit. The spectral functions, shown in panel~(a), have a similar form, except for the additional sharp central peak associated with the diffusion of vacancies in the crystal with vacancies. This additional peak becomes a long-time decaying mode in the intermediate scattering function with an amplitude given by the factor $D_{\rm l}$. Indeed, the two intermediate scattering functions, shown in panel~(b), have a similar damped oscillating structure at short times, due to the diffusion of heat and the longitudinal propagation of sound. On intermediate-time scales, they differ by a factor of order $D_{\rm l}$, providing a clear signature of the eighth mode of vacancy diffusion. On very long-time scales, both functions asymptote to zero.

%%%%%%%%%%%%%%%%%%%%%%%%%%%%%%%%%%%%%%%%%%%%%%%%%%%%%%%%%%%%%%%%
%%%%%%%%%%%%%%%%%%%%%%%%%%%%%%%%%%%%%%%%%%%%%%%%%%%%%%%%%%%%%%%%

%%%%%%%%%%%%%%%%%%%%%%%%%%%%%%%%%%%%%%%%%%%%%%%%%
\section{Hard-sphere crystal}
\label{sec:MDHS}
%%%%%%%%%%%%%%%%%%%%%%%%%%%%%%%%%%%%%%%%%%%%%%%%%

The intermediate scattering function of a monatomic face-centered cubic hard-sphere crystal with one vacancy is computed using a molecular dynamics simulation. The result is compared with the Fourier transform of the theoretical expression for the dynamic structure factor given in Eq.~\eqref{eq:DSF_Full}, and with the ISF of a perfect crystal.  Details on the simulation are given in Appendix~\ref{app:MDHS}.

\subsection{Results of the simulation}

%%%%%%%%%%%%%%%%%%%%%%%%%%%%%%%%%%%%%%%%%%%%%%%%%%%%%%%%%%%%%%%%
\begin{figure}[t!] \centering
    \begin{minipage}{0.48\textwidth}
        \centering
        \includegraphics[width=1.\textwidth]{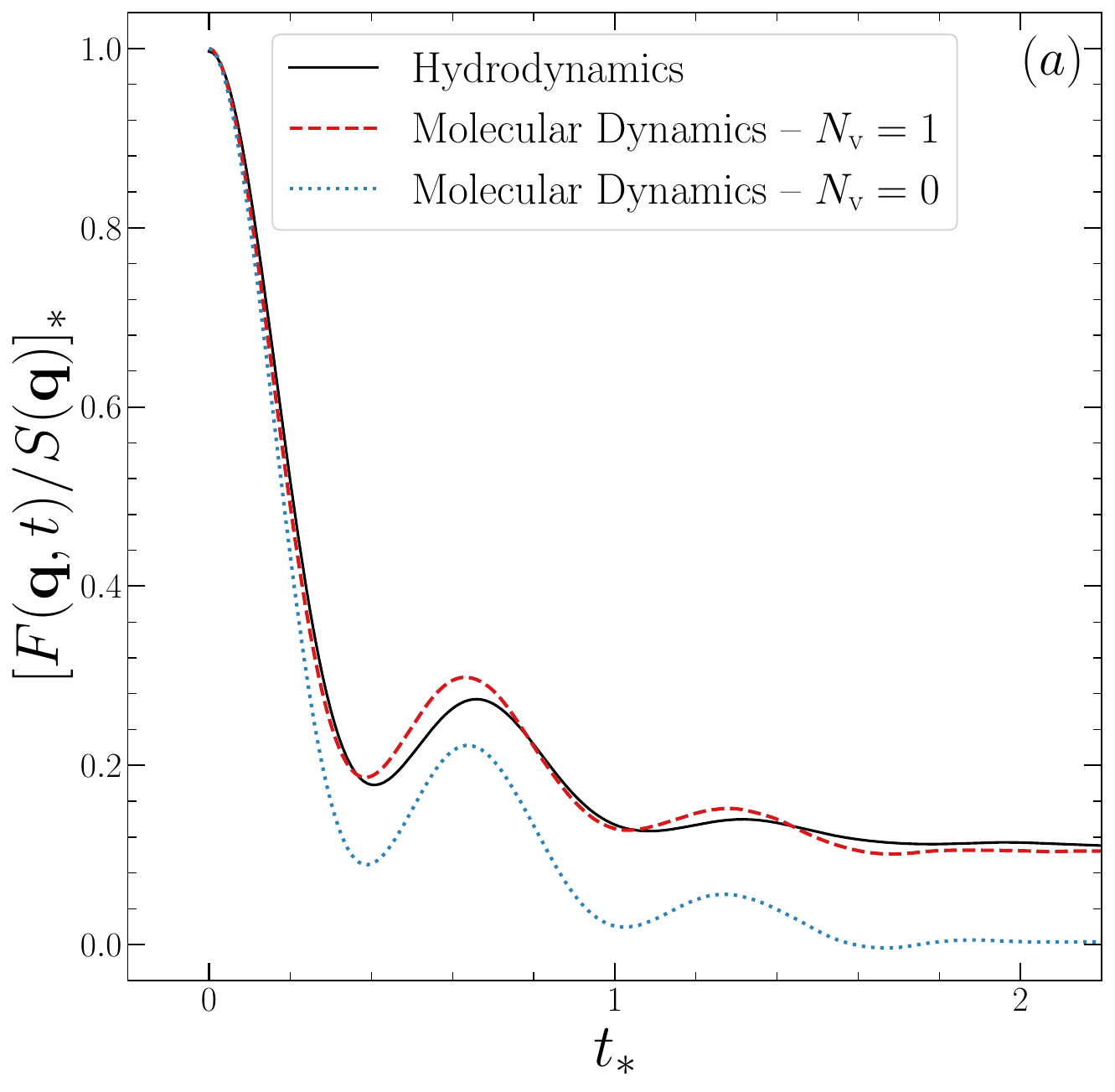} % second figure itself
    \end{minipage}\hfill
     \begin{minipage}{0.48\textwidth}
        \centering
        \includegraphics[width=1.\textwidth]{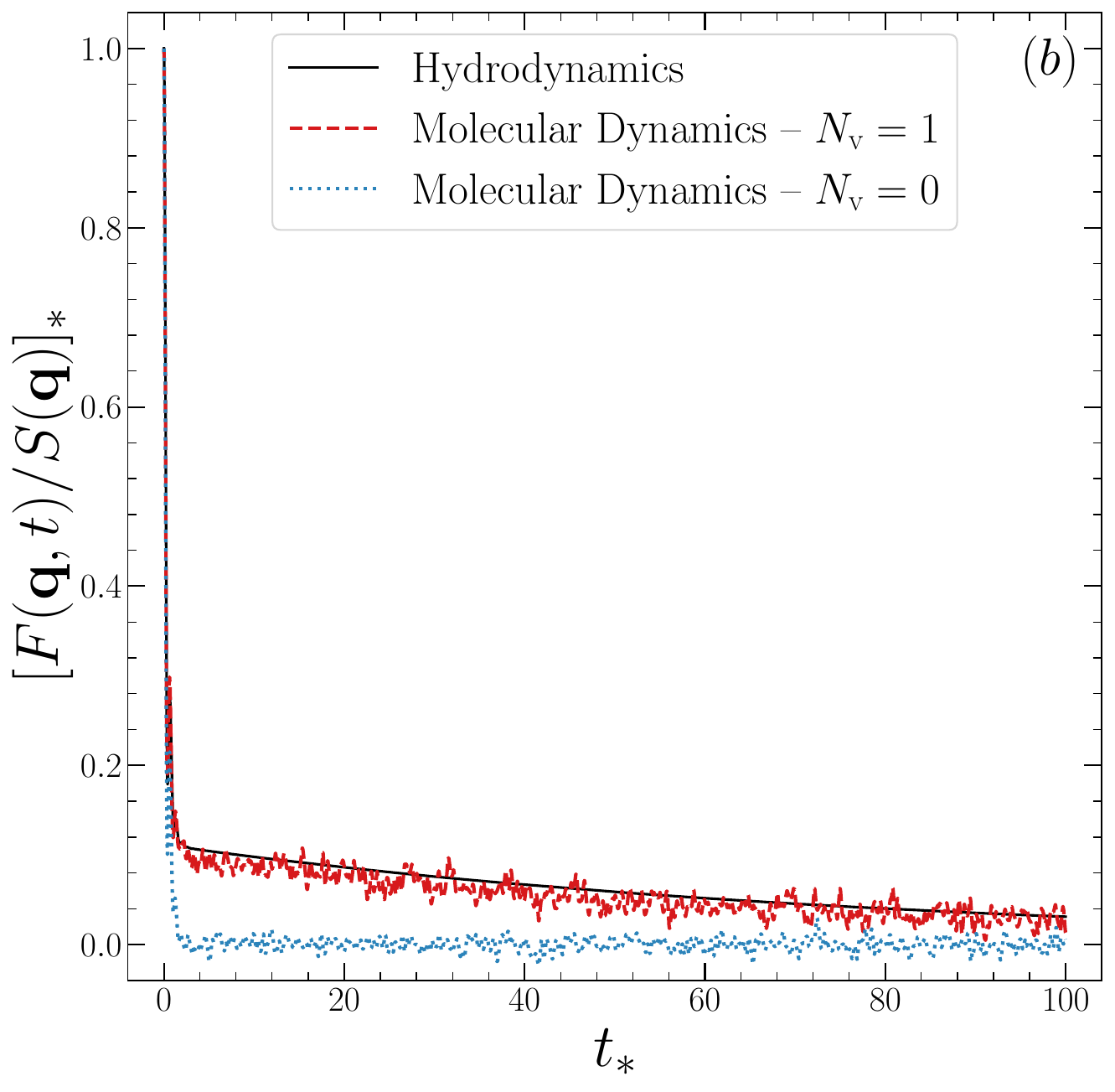} % first figure itself
    \end{minipage}
\caption[] {Normalized intermediate scattering function $F({\bf q},t)$ at density  $n_{0*}=1.05$, with ${\bf q}_*$ in the $[100]$ direction and $q_*=0.8$, for a hard-sphere crystal with $N_0=500$ on short, panel~(a), and long, panel~(b),  time scales. The dashed red line corresponds to a crystal with one vacancy and 499 particles. The dotted blue line corresponds to a perfect crystal with 500 particles.  The solid black line is the inverse Fourier transform from frequency to time of the analytical expression~\eqref{eq:DSF_Full}, with parameters given in Ref.~\cite{MG25} for the thermodynamic, elastic, and transport coefficients, and in Appendix~\ref{app:static-fns} for the static correlation functions. } \label{Fig:ISF_MDHS}
\end{figure}
%%%%%%%%%%%%%%%%%%%%%%%%%%%%%%%%%%%%%%%%%%%%%%%%%%%%%%%%%%%%%%%%

The system is at a lattice density of $n_{0*}=1.05$ (in reduced units, see Appendix~\ref{app:MDHS}), with a wave vector in the direction $[100]$. We chose a density slightly above the melting density, $n_{\rm m*}= 1.037$, because the vacancy diffusion coefficient decreases with increasing density. This choice ensures the  stability of the crystal while allowing for a sufficient number of vacancy jumps to be simulated. Our theoretical framework applies in the hydrodynamic limit, which requires a small wave number $q$, typically $q_*<1$. Moreover, the coefficient $D_{\rm l}$, Eq.~\eqref{eq:ISF}, is linear in the molar fraction of vacancies $y$. Finally, to observe the long-time decay of the vacancy diffusion mode, it is necessary to compute the ISF over a long-time interval. Based on these requirements and our available computational resources, we simulate a crystal of $500$ lattice sites with one vacancy. The wave number $q_*=0.8$ is in the hydrodynamic regime, and the molar fraction of vacancies, $y=2\times10^{-3}$, is such that the decaying vacancy diffusion mode is distinguishable from  numerical noise. Furthermore, the moderate crystal size allows us to  compute  the ISF over a long enough time interval. The result of the simulation is shown in Fig.~\ref{Fig:ISF_MDHS}.

On short-times scales, as seen in panel~(a), the result of the molecular simulation (dashed red line) is in excellent agreement with the theoretical prediction (full black line) obtained from the numerical inverse Fourier transform of Eq.~\eqref{eq:DSF_Full}. The small deviations, observed for example around the first local maximum of the oscillation, are explained by the relatively high value of the wave number, preventing the simulation to fully reach the hydrodynamic regime where the macroscopic theory is expected to apply. At initial times, the decay of the correlation function is dominated by the acoustic and heat modes, and the oscillations stem from the  propagation of  sound waves. The decay and oscillations are also observed for the perfect crystal (dotted blue line). At these early times, the behavior of the ISF is similar for both perfect and  nonperfect crystals. However, while the former decays to zero over the time interval considered in panel~(a), the latter is still finite  with a value that is compatible with the prefactor that includes the coefficient~$D_{\rm l}$. This is a clear signature of the vacancy diffusion mode and supports the results in Eqs.~\eqref{eq:ISF}-\eqref{eq:coeffDofy}.

To numerically study the decay of the ISF for a crystal with vacancies, the function is computed over long-time scales, as shown in panel~(b). The result of the molecular simulation (dashed red line) is again in excellent agreement with the theoretical prediction (full black line) obtained from the inverse Fourier transform of Eq.~\eqref{eq:DSF_Full}. The theory predicts that at sufficiently long times, the heat and acoustic modes are suppressed and the ISF is dominated by the vacancy diffusion mode. The numerical simulation indeed shows a contribution consistent with vacancy diffusion.  Moreover, the decay is not linear, but compatible with an exponential decay in time given by ${\rm e}^{-{\cal D}_yq^2t}$. The vacancy diffusion coefficient ${\cal D}_y$ can be determined from the decay of the intermediate scattering function, which could be computationally more efficient than tracking the position of the vacancy throughout the simulation.  After a long enough time, the result for the ISF of the perfect crystal (dotted blue line) shows numerical fluctuations around zero, since, in the absence of the eighth mode, the function has already decayed.

The parameters used in the theoretical expression are the thermodynamic, elastic, and transport coefficients appearing in Eqs.~\eqref{eq:D1}-\eqref{eq:M2}. These coefficients  are derived from microscopic expressions based on the statistical-mechanical theory of transport in crystals~\cite{MG21,MG25}, and are computed for a hard-sphere crystal in Refs.~\cite{MG24a,MG25}. The excellent agreement shown in Fig.~\ref{Fig:ISF_MDHS} between the theoretical prediction using these coefficients as parameters, and the direct numerical computation of the DSF provides evidence for the statistical-mechanical theory of transport in crystals.

 Finally, note that we do not compute the dynamic structure factor numerically. To do so, it would be necessary to obtain the ISF for a time interval over which it has completely decayed before taking a numerical Fourier transform. Such a simulation would exceed our  computational resources. Nevertheless, the ISF alone is sufficient to investigate all the theoretical predictions presented here.

%%%%%%%%%%%%%%%%%%%%%%%%%%%%%%%%%%%%%%%%%%%%%%%%%
\section{Conclusion and perspectives}
\label{sec:conclusion}
%%%%%%%%%%%%%%%%%%%%%%%%%%%%%%%%%%%%%%%%%%%%%%%%%

This work presents a study of the dynamic structure factor of a monatomic cubic crystal. Working in the three special directions of the wave vector, where the eight macroscopic hydrodynamic equations describing transport in one-component crystals split into longitudinal and transverse sets, we have derived in Eq.~\eqref{eq:DSF_Full} an exact analytical expression for the DSF, which is the central result of this work. From this formal expression, the resonances are identified and interpreted. The poles of the spectral function give the dispersion relations~\eqref{eq:z_H}-\eqref{eq:z_pm}, which agree with earlier works~\cite{FC76,MG21,MG25}.  An expansion for  small values of the wave number gives the approximate expression~\eqref{eq:DSF_approx} for the DSF and the corresponding intermediate scattering function~\eqref{eq:ISF}. The approximate and full expressions  agree  in the hydrodynamic limit. From the approximate expression, the resonances are characterized as Lorentzian functions, including a Rayleigh central peak for heat diffusion, a Brillouin doublet for longitudinal sound propagation and an additional central peak corresponding to the diffusion of vacancies. These results generalize earlier works on the spectral functions of the perfect crystal~\cite{MG24b} and the fluid~\cite{F75,BP76,BY80,M66}.  We also obtain analytical expressions, with explicit contributions stemming from the presence of vacancies, for the speeds of sound [Eqs.~\eqref{eq:tcs} and~\eqref{eq:lcs}], the sound damping coefficients [Eqs.~\eqref{eq:Gammatfull} and~\eqref{eq:Gammalfull}], and the heat and vacancy diffusivities [Eqs.~\eqref{eq:chifull} and~\eqref{eq:Dvacfull}]. These expressions are another key result of this work and play a central role in the study of transport phenomena in crystals, both theoretically and numerically, and for connecting with experiments.

The theoretical results from Sec.~\ref{sec:DSF} are confirmed by molecular dynamics simulations. In Sec.~\ref{sec:MDHS}, we  compute the intermediate scattering function for a fcc hard-sphere crystal with one vacancy. The results agree with the predictions of hydrodynamic theory across small, intermediate, and long-time scales. On small-time scales, we observe the oscillatory behavior from the longitudinal sound propagation, and the decay of the function from the heat and acoustic modes. On intermediate-time scales, the heat and acoustic modes are suppressed, and the ISF is dominated by the mode associated with vacancy diffusion. On long-time scales, the   behavior of the ISF is consistent with the exponential decay of the vacancy diffusion mode. The coefficient  ${\cal D}_y$ of vacancy diffusion can thus be determined from the decay of the ISF, without explicitly tracking the vacancies positions during the simulation. The presence of this eighth slow mode gives the ISF a characteristic signature, distinguishing it from that of a perfect crystal.

The theoretical and numerical results presented here are for a monatomic cubic crystal, with the wave vector taken in the special directions $[100]$, $[110]$, and $[111]$. For other directions, the system of closed equations~\eqref{eq:deltay}-\eqref{eq:deltauab} does not decouple. In this case, the DSF would contain eight distinct resonances, including contributions from the transverse modes. In non-cubic crystals, the set of closed equations is not expected to decouple either. Furthermore, for the $20$ crystallographic classes compatible with piezoelectricity, the transport equations~\eqref{eq:drho}-\eqref{eq:du} include additional terms with rank-three tensors. These terms describe cross-effects that couple momentum  to heat or vacancy transport. As demonstrated in Ref.~\cite{MG21}, the presence of these tensors breaks the degeneracy of the damping rates for transverse sound modes propagating in opposite directions. This effect is expected to break the even symmetry of the DSF with respect to frequency. Finally, for multicomponent crystals, which include different species of particles like atoms, molecules, or colloids, there is an additional slow mode associated with each  supplementary species, and  ruling mutual diffusion between them. This leads to additional resonances in the DSF.  

Numerical studies of  realistic systems may require moving beyond the hard-sphere potential and event-driven algorithms. Lennard-Jones potentials could serve as an example, as they are used for simulations of rare-gas crystals. These more advanced approaches are necessary to accurately model real-world crystals and their experimental behaviors. These investigations are left for future work.

A key finding of this work is the identification and characterization of the resonance  linked to vacancy diffusion within the crystal, which appears as a sharp central peak in the dynamic structure factor with a width given by the vacancy diffusivity and a prefactor $D_{\rm l}$, given in Eq.~\eqref{eq:coeffDofy}. This result provides further evidence for the phenomenological theory, conjectured in the 1970s~\cite{MPP72,FC76}, that vacancy diffusion is indeed the eighth slow mode of a monatomic crystal. Our work completes and confirms the recent microscopic approach presented in Ref.~\cite{MG25} by providing an alternative proof that ${\cal D}_y$ is indeed the vacancy diffusion coefficient. In Ref.~\cite{MG25}, the vacancy spectral function $\left\langle \delta y({\bf q},\omega)\delta y^*({\bf q},0)\right\rangle_{\rm eq}$ was studied numerically. Using an approach similar to the one presented here for the DSF, an analytical expression for  this function could be obtained. There is also the question of several vacancies, still in the dilute limit,  interacting and presenting collective dynamics. Our theory applies in this context, but the order ${\cal O}(y)$ corrections of all quantities should be included. On the numerical side, the dilute limit with multiple vacancies necessitates the consideration of much larger crystals. We leave these prospects open for future studies.

In conclusion, this work  provides new insights into transport phenomena in crystals by building upon a recently developed statistical-mechanical theory of dissipative hydrodynamics for solids. The study provides further evidence for this approach by validating the transport coefficients, which were previously computed  using microscopic formulas. Our research represents a step forward, moving from idealized models to the study of real-world crystals, and contributes to a deeper understanding of the fundamental mechanisms that govern the collective behavior of matter in the solid state.

%%%%%%%%%%%%%%%%%%%%%%%%%%%%%%%%%%%%%%%%%%%%%%%%%
\section*{Acknowledgements}
%%%%%%%%%%%%%%%%%%%%%%%%%%%%%%%%%%%%%%%%%%%%%%%%%
A. Yerle and J. Mabillard thank the Université Paris Cité and the Laboratoire Matière et Systèmes Complexes (M.S.C.) for support in this research.  P. Gaspard thanks the Universit\'e Libre de Bruxelles (U.L.B.) for support in this research.

%%%%%%%%%%%%%%%%%%%%%%%%%%%%%%%%%%%%%%%%%%%%%%%%%
\appendix
%%%%%%%%%%%%%%%%%%%%%%%%%%%%%%%%%%%%%%%%%%%%%%%%%

\section{Closing the transport equations with crystal thermodynamics}
\label{app:TH}

We detail the derivation of the closed set of equations~\eqref{eq:deltay}-\eqref{eq:deltauab} from the transport equations~\eqref{eq:drho}-\eqref{eq:du}, which originate from the statistical-mechanical theory of transport in crystals. Following a methodology similar to that presented in Ref.~\cite{MG25}, we use relations from crystal thermodynamics~\cite{W98}.

The fluctuations of the entropy per unit mass (or specific entropy)~${\mathfrak s}\equiv s/\rho$ are given in terms of $\delta \epsilon_0$ and $ \delta \rho$ as
\begin{align}
\delta {\mathfrak s} = \frac{1}{\rho T} \left( \delta \epsilon_0 - \frac{\epsilon_0+p}{\rho}\, \delta \rho\right)\;.
\end{align}
Given the evolution equations~\eqref{eq:drho} for $ \delta \rho$ and~\eqref{eq:de} for $ \delta \epsilon_0$, the evolution equation for $\delta {\mathfrak s}$ is derived as
\begin{align}
	\rho T \partial_t\delta {\mathfrak s}&= \kappa^{\prime} \, \nabla^2 \delta T + \xi \, \nabla^a \nabla^b \delta \phi^{ab} - \xi \, \mu_{y,\boldsymbol{\phi}} \, \nabla^2\delta\phi^{aa} - {\cal K}_y \, \nabla^2 \delta y \, .\label{eq:ds} 
\end{align}
The evolution equation for the fluctuations of the strain tensor $\delta u^{ab}$ is derived from the evolution equation~\eqref{eq:du} for the displacement field $\delta u^a$ as
\begin{align}
	\partial_t \, \delta u^{ab} & = \frac{1}{2}\left(\nabla^a\delta v^b+\nabla^b\delta v^a\right) +\frac{\xi^{\prime}}{T} \,  \nabla^a\nabla^b \delta T + \frac{\zeta}{2} \, \nabla^c\left( \nabla^a \delta \phi^{cb}+ \nabla^b \delta \phi^{ca}\right) - \zeta \, \mu_{y,\boldsymbol{\phi}} \,  \nabla^a\nabla^b\delta \phi^{cc}- {\cal D}_{y} \,  \nabla^a\nabla^b\delta y \, . \label{eq:duab}
\end{align}

To close the system,  the quantities $(\delta {\mathfrak s}, \delta\sigma^{ab},\delta\phi^{ab})$ are expressed in terms of the variables $(\delta y, \delta T, \delta u^{ab})$ as 
\begin{align}
\delta {\mathfrak s} &= \frac{\alpha\, B_T}{\rho}\, \delta u^{aa} + \frac{c_v}{T}\, \delta T + \varsigma_y \, \delta y \, , \label{s-uTy-cubic}\\
\delta \sigma^{ab} &= B_T^{abcd} \,  \delta u^{cd} - \alpha B_T \, \delta^{ab} \, \delta T - \pi_y \, \delta^{ab} \, \delta y \, , \label{sigma-uTy-cubic}\\
\delta \phi^{ab} &= G_T^{abcd} \, \delta u^{cd} \, , \label{phi-uTy-cubic}
\end{align}
using the results of Ref.~\cite{MG25}. Substituting equations~\eqref{s-uTy-cubic}-\eqref{phi-uTy-cubic} into the transport equations~\eqref{eq:drhov},~\eqref{eq:dy},~\eqref{eq:ds} and~\eqref{eq:duab} gives the closed system~\eqref{eq:deltay}-\eqref{eq:deltauab}. This is achieved by using the relation $G_T^{aacd} = 0$, which holds for cubic crystals, and the symmetries  $\eta^{abcd}=\eta^{bacd}$, $B_T^{abcd}=B_T^{bacd}$, and $G_T^{abcd}=G_T^{bacd}$.

%%%%%%%%%%%%%%%%%%%%%%%%%%%%%%%%%%%%%%%%%%%%%%%%%%%%%%%%%%
\section{Fourier-Laplace transform and separation into transverse and longitudinal equations}
\label{app:FLTandLT}
%%%%%%%%%%%%%%%%%%%%%%%%%%%%%%%%%%%%%%%%%%%%%%%%%%%%%%%%%%

We obtain the sets of longitudinal and transverse evolution equations~\eqref{eq:lrho}-\eqref{eq:tu} in Fourier-Laplace space.
%%%%%%%%%%%%%%%%%%%%%%%%%%%%%%%%%%%%%%%%%%%%%%%%%%%%%%%%%%
\subsection{Fourier-Laplace transform of the closed equations}
\label{app:FLT}

Taking the Fourier-Laplace transform~\eqref{eq:FLT} of the set of closed equations~\eqref{eq:deltay}-\eqref{eq:deltauab} gives
\begin{align}
&\left(z+q^2 {\cal D}_y\right)\delta \tilde{y}({\bf q},z)-q^2\frac{\xi'}{T}\delta \tilde{T}({\bf q},z)-q^aq^b G_T^{abcd}  \zeta   \delta {\tilde{u}}^{cd}({\bf q},z)=\delta {y}({\bf q},0)\;,\label{eq:FLdeltay}\\
&\left(z\frac{\alpha B_T}{\rho}\delta^{cd}+ q^aq^b\frac{G_T^{abcd}  \xi}{\rho T}\right)  \delta \tilde{u}^{cd}({\bf q},z) +\left(z \frac{c_v}{T}+q^2 \frac{ \kappa^{\prime} }{\rho T}\right) \delta \tilde{T}({\bf q},z) + \left(\varsigma_y z-q^2 \frac{ {\cal K}_y}{\rho T}  \right)\delta \tilde{y}({\bf q},z) \notag\\& \qquad= \frac{\alpha B_T}{\rho} \delta {u}^{aa}({\bf q},0)+ \frac{c_v}{T}\delta {T}({\bf q},0)+\varsigma_y\delta{y}({\bf q},0)\; ,\label{eq:FLdeltaT}\\
&\left(z\delta^{ad}+ q^bq^c\frac{\eta^{abcd}}{\rho} \right)\delta \tilde{v}^d({\bf q},z)-{\rm i}q^a\frac{\pi_y}{\rho}\delta\tilde{y}({\bf q},z)-{\rm i}q^a  \frac{\alpha B_T }{\rho }  \delta \tilde{T}({\bf q},z)+{\rm i}q^b \frac{B_T^{abcd}}{\rho}\delta \tilde{u}^{cd}({\bf q},z)=\delta {v}^a({\bf q},0) \;,\label{eq:FLdeltav}\\
&\left(z\delta^{ae}\delta^{bf}+q^bq^d\frac{G_T^{adef}  \zeta}{2}+q^aq^d\frac{G_T^{bdef}  \zeta}{2} \right)  \delta \tilde{u}^{ef}({\bf q},z)-q^aq^b{\cal D}_y{\delta} \tilde y({\bf q},z)+q^aq^b\frac{ \xi^{\prime}}{T}\delta \tilde{T}({\bf q},z)+\frac{1}{2}\left[ {\rm i}q^a\delta \tilde{v}^b({\bf q},z)+{\rm i}q^b\delta \tilde{v}^a({\bf q},z)\right]\notag\\ &\qquad =\delta {u}^{ab}({\bf q},0)\,.\label{eq:FLdeltauab}
\end{align}

%%%%%%%%%%%%%%%%%%%%%%%%%%%%%%%%%%%%%%%%%%%%%%%%%%%%%%%%%%
\subsection{Transverse and longitudinal equations}
\label{app:TLE}

In the orthonormal basis $\{{\bf e}_{\rm l},{\bf e}_{\rm t_1},{\bf e}_{\rm t_2} \}$, we have
\begin{align}
	{\bf q}=q{\bf e}_{\rm l}\;,	&&	{\bf v}=v_\sigma{\bf e}_\sigma = v_{\rm l}{\bf e}_{\rm l}+v_{{\rm t}_k}{\bf e}_{{\rm t}_k}\;, && {\bf u}=u_\sigma{\bf e}_\sigma = u_{\rm l}{\bf e}_{\rm l}+u_{{\rm t}_k}{\bf e}_{{\rm t}_k}\;,
\end{align}
where $\sigma\in\{{\rm l,t_1,t_2}\}$. The strain tensor becomes
\begin{align}
	u^{ab}({\bf q})=-\frac{{\rm i}}{2}\left[q^au^b({\bf q})+q^bu^a({\bf q})\right]=-\frac{{\rm i}}{2}qu_\sigma({\bf q})\left(e_{\rm l}^ae_\sigma^b+e_{\rm l}^be_\sigma^a\right)\,. \label{eq:uabqapp}
\end{align}
 The longitudinal component is computed as
\begin{align}
	u^{ab}({\bf q})e^a_{\rm l}e^b_{\rm l}&=-{\rm i}qu_\sigma({\bf q})({\bf e}_{\rm l}\cdot {\bf e}_{\rm l})({\bf e}_\sigma\cdot{\bf e}_{\rm l})=-{\rm i}qu_\sigma({\bf q})\delta_{\rm l \sigma}\;,\label{eq:appulq}
\end{align}
implying $u_{\rm l}({\bf q})={\rm i}q^{-1}u^{ab}({\bf q})e^a_{\rm l}e^b_{\rm l}$. Similarly, for the transverse components, we have
\begin{align}
	u^{ab}({\bf q})e^a_{\rm l}e^b_{{\rm t}_k}&=-\frac{{\rm i}}{2}qu_\sigma({\bf q})\left[({\bf e}_{\rm l}\cdot {\bf e}_{\rm l})({\bf e}_\sigma\cdot{\bf e}_{{\rm t}_k})+({\bf e}_{\rm l}\cdot {\bf e}_{{\rm t}_k})({\bf e}_\sigma\cdot{\bf e}_{\rm l})\right]=-\frac{{\rm i}}{2}qu_\sigma({\bf q})\delta_{\sigma {\rm t}_k}\;,\label{eq:apputq}
\end{align}
giving $u_{{\rm t}_k}({\bf q})=2{\rm i}q^{-1}u^{ab}({\bf q})e^a_{\rm l}e^b_{{\rm t}_k}$.
From Eq.~\eqref{eq:uabq}, we have
\begin{align}
	u^{aa}({\bf q})&=-{\rm i}qu_\sigma({\bf q}){\bf e}_{\rm l}\cdot{\bf e}_\sigma=-{\rm i}qu_\sigma({\bf q})\delta_{\rm l\sigma}=-{\rm i}qu_{\rm l}({\bf q})\;.\label{eq:appuaaq}
\end{align}
Having expressed the vector and tensorial quantities in the basis $\{{\bf e}_{\rm l},{\bf e}_{\rm t_1},{\bf e}_{\rm t_2} \}$, we now split the set~\eqref{eq:FLdeltay}-\eqref{eq:FLdeltauab} into longitudinal and transverse equations.

\subsubsection{The longitudinal equation~\eqref{eq:ly} for $\delta \tilde y$} 

The term involving $\delta \tilde u^{cd}$ in Eq.~\eqref{eq:FLdeltay} is simplified as
\begin{align}
q^aq^b G_T^{abcd}  \zeta   \delta {\tilde{u}}^{cd}({\bf q},z)&=-{\rm i}q^3\delta \tilde u_\sigma({\bf q},z)\zeta G_T^{abcd}e_{\rm l}^ae_{\rm l}^ce_{\rm l}^be_\sigma^d\notag\\
&=q^2[-{\rm i}q\delta \tilde u_{\rm l}({\bf q},z)]\zeta G_{T,{\rm l}}\notag\\
&=-q^2\left[\delta \tilde y({\bf q})+\frac{\delta\tilde\rho({\bf q})}{\rho}\right]\zeta G_{T,{\rm l}}\;,\label{eq:uabtermy}
\end{align}
using the symmetry $G_T^{abcd}=G_T^{abdc}$, and $G_T^{abcd}e_{\rm l}^ae_{\rm l}^ce_{\rm l}^be_\sigma^d=G_{T,{\rm l}\sigma}=G_{T,{\rm l}}\delta_{{\rm l}\sigma}$.  The latter follows from Eqs.~\eqref{eq:4tbasis} and~\eqref{eq:2tbasis},  which hold when ${\bf q}$ is along the special directions. Substituting Eq.~\eqref{eq:uabtermy} into Eq.~\eqref{eq:FLdeltay} yields Eq.~\eqref{eq:ly}.

\subsubsection{The longitudinal equation~\eqref{eq:lT} for $  \delta \tilde T$} 

Following identical steps as above, the term involving $G_{T}^{abcd}$ in Eq.~\eqref{eq:FLdeltaT}  gives
\begin{align}
	\frac{G_T^{abcd}  \xi}{\rho T}q^aq^b  \delta \tilde{u}^{cd}({\bf q},z)&=-\frac{\xi G_{T,{\rm l}}}{\rho T}q^2\left[\delta \tilde y({\bf q})+\frac{\delta\tilde\rho({\bf q})}{\rho}\right]\;.
\end{align}
For the other terms involving $\delta\tilde u^{aa}$, we use the relations~\eqref{eq:ul} and~\eqref{eq:appuaaq}.  Eq.~\eqref{eq:FLdeltaT} is longitudinal and reads
\begin{align}
&-\left(z\frac{\alpha B_T}{\rho^2}+ q^2\frac{G_{T,{\rm l}} \xi}{\rho^2 T}\right)  {\delta\tilde\rho({\bf q},z)} +\left(z \frac{c_v}{T}+ q^2\frac{ \kappa^{\prime} }{\rho T}\right) \delta \tilde{T}({\bf q},z) + \left[z\left(\varsigma_y-\frac{\alpha B_T}{\rho}\right)- q^2\left( \frac{G_{T,{\rm l}} \xi}{\rho T}+\frac{ {\cal K}_y}{\rho T} \right) \right]\delta \tilde{y}({\bf q},z) \notag\\
& \qquad= -\frac{\alpha B_T}{\rho^2}\delta\rho({\bf q},0)+ \frac{c_v}{T}\delta {T}({\bf q},0)+\left(\varsigma_y-\frac{\alpha B_T}{\rho}\right)\delta{y}({\bf q},0) \;.\label{eq:lT0}
\end{align}

\subsubsection{The longitudinal and transverse equations~\eqref{eq:lv} and~\eqref{eq:tv}  for $\delta \tilde v^a$}

The equation~\eqref{eq:FLdeltav} for the three components of the velocity field is split into one longitudinal and two transverse equations. In the orthonormal basis, the terms of Eq.~\eqref{eq:FLdeltav} are
\begin{align}
	z\delta \tilde{v}^a({\bf q},z)&=z \delta \tilde v_{\sigma}({\bf q},z){e}^a_{\sigma}\,,\\
	 \frac{\eta^{abcd}}{\rho} q^bq^c\delta \tilde{v}^d({\bf q},z) &=\frac{q^2}{\rho} \delta \tilde v_{\sigma}({\bf q},z)\eta^{abcd}e_{\rm l}^be_{\rm l}^c{e}^d_{\sigma}\;,\\
	 -{\rm i}q^a\frac{\pi_y}{\rho}\delta\tilde{y}({\bf q},z)&=-{\rm i}qe_{\rm l}^a\frac{\pi_y}{\rho}\delta\tilde{y}({\bf q},z)\,,\\
	 -{\rm i}q^a  \frac{\alpha B_T }{\rho }  \delta \tilde{T}({\bf q},z)&=-{\rm i}qe_{\rm l}^a  \frac{\alpha B_T }{\rho }  \delta \tilde{T}({\bf q},z)\;,\\
	 {\rm i}q^b \frac{B_T^{abcd}}{\rho}\delta \tilde{u}^{cd}({\bf q},z)&=\frac{q^2}{\rho}\delta \tilde u_\sigma({\bf q})B_T^{abcd}e_{\rm l}^be_{\rm l}^ce_\sigma^d\;,
\end{align}
using  the symmetry $B_T^{abcd}=B_T^{abdc}$ for the last equality. 

Taking the scalar product of Eq.~\eqref{eq:FLdeltav} with ${\bf e}_{\rm l}$ gives
\begin{align}
	\left(z+ q^2\frac{\eta_{\rm l}}{\rho}\right)\delta \tilde v_{\rm l}({\bf q},z)-{\rm i}q\frac{\pi_y}{\rho}\delta\tilde{y}({\bf q},z)-{\rm i}q  \frac{\alpha B_T }{\rho }  \delta \tilde{T}({\bf q},z)+\frac{q^2}{\rho} B_{T,{\rm l}}\delta \tilde u_{\rm l}({\bf q},z)=\delta  v_{\rm l}({\bf q},0)\;,
\end{align}
where we have introduced the rank-two tensors from Eq.~\eqref{eq:4tbasis}, which are diagonal for ${\bf q}$ along the special directions according to Eq.~\eqref{eq:2tbasis}. Substituting Eq.~\eqref{eq:ul} gives the longitudinal equation~\eqref{eq:lv} for the velocity.

Finally, taking the scalar product of Eq.~\eqref{eq:FLdeltav} with ${\bf e}_{{\rm t}_k}$, for $k=1,2$, yields the transverse equation~\eqref{eq:tv}.

\subsubsection{The longitudinal and transverse equations~\eqref{eq:lrho} and~\eqref{eq:tu} for $\delta\tilde u^{ab}$}

The equation~\eqref{eq:FLdeltauab} for the components of the strain tensor is split into one longitudinal and two transverse equations. In the orthogonal basis, the terms of Eq.~\eqref{eq:FLdeltauab} are
\begin{align}
	\frac{q^bq^d}{2}G_T^{adef}  \zeta \delta \tilde{u}^{ef}({\bf q},z)&=-{\rm i}\frac{\zeta}{2}q^3\delta\tilde u_\sigma({\bf q},z)G_T^{adef}e_{\rm l}^be_{\rm l}^de_{\rm l}^ee_\sigma^f\;,\\
	q^aq^b{\cal D}_y{\delta} \tilde y({\bf q},z)&=q^2e_{\rm l}^ae_{\rm l}^b{\cal D}_y{\delta} \tilde y({\bf q},z)\;,\\
	\frac{ \xi^{\prime}}{T}q^aq^b\delta \tilde{T}({\bf q},z)&=\frac{ \xi^{\prime}}{T}q^2 e_{\rm l}^ae_{\rm l}^b\delta \tilde{T}({\bf q},z)\;,\\
	\frac{i}{2}q^a\delta \tilde{v}^b({\bf q},z)&=\frac{iq}{2}\delta\tilde v_\sigma({\bf q},z)e_{\rm l}^ae_\sigma^b\;,
\end{align}
using the symmetry $G_T^{abcd}=G_T^{abdc}$.

Taking the double scalar product of Eq.~\eqref{eq:FLdeltauab} with ${{\bf e}_{\rm l}}$ and ${\bf e}_{\rm l}$ gives
\begin{align}
&e_{\rm l}^ae_{{\rm l}}^b\left\{\left[ z \delta \tilde{u}^{ab}({\bf q},z)-{\rm i}\frac{\zeta}{2}q^3 \delta \tilde u_\sigma({\bf q},z)G_T^{adef}e_{\rm l}^be_{\rm l}^de_{\rm l}^ee_\sigma^f-{\rm i}\frac{\zeta}{2}q^3 \delta \tilde u_\sigma({\bf q},z)G_T^{bdef}e_{\rm l}^ae_{\rm l}^de_{\rm l}^ee_\sigma^f\right] -q^aq^b{\cal D}_y{\delta} \tilde y({\bf q},z)\right.\notag\\
&\left.+\frac{ \xi^{\prime}}{T}q^aq^b\delta \tilde{T}({\bf q},z)+\frac{{\rm i}q}{2} \delta \tilde v_\sigma({\bf q},z)\left[ e_{\rm l}^ae_\sigma^b+ e_{\rm l}^be_\sigma^a\right]\right\}=e_{\rm l}^ae_{{\rm t}_k}^b\delta {u}^{ab}({\bf q},0)\;.
\end{align}
Using similar steps as above, we find
\begin{align}
	e_{\rm l}^ae_{\rm l}^b\delta\tilde u^{ab}({\bf q},z)&=-\delta\tilde y({\bf q},z)-\frac{\delta\tilde\rho({\bf q},z)}{\rho}\;,\\
	e_{\rm l}^ae_{\rm l}^b\left[-{\rm i}\frac{\zeta}{2}q^3\delta\tilde u_\sigma({\bf q},z)G_T^{adef}e_{\rm l}^be_{\rm l}^de_{\rm l}^ee_\sigma^f\right]&=-\frac{\zeta}{2} G_{T,{\rm l}}q^2\left[\delta\tilde y({\bf q},z)+\frac{\delta\tilde\rho({\bf q},z)}{\rho}\right]\;,\\
	e_{\rm l}^ae_{\rm l}^bq^2e_{\rm l}^ae_{\rm l}^b{\cal D}_y{\delta} \tilde y({\bf q},z)&=q^2{\cal D}_y{\delta} \tilde y({\bf q},z)\;,\\
	e_{\rm l}^ae_{\rm l}^b\frac{ \xi^{\prime}}{T}q^2 e_{\rm l}^ae_{\rm l}^b\delta \tilde{T}({\bf q},z)&=\frac{ \xi^{\prime}}{T}q^2 \delta \tilde{T}({\bf q},z)\;,\\
	e_{\rm l}^ae_{\rm l}^b\left[{\rm i}q\delta\tilde v_\sigma({\bf q},z)e_{\rm l}^ae_\sigma^b\right]&={\rm i}q\delta\tilde v_{\rm l}({\bf q},z)\;.
\end{align}
Putting everything together, the longitudinal equation stemming  from $\delta \tilde u^{ab}$ is found as
\begin{align}
	-\left[z+q^2\left({\cal D}_y+{\zeta G_{T,{\rm l}}}\right)\right]\delta \tilde y({\bf q},z)-\left(z+q^2{\zeta G_{T,{\rm l}}}\right)\frac{\delta\tilde\rho({\bf q},z)}{\rho}+q^2\frac{ \xi^{\prime}}{T} \delta \tilde{T}({\bf q},z)+{\rm i}q\delta \tilde  v_{\rm l}({\bf q},z)& =-\delta  y({\bf q},0)-\frac{\delta\rho({\bf q},0)}{\rho}\;.\label{eq:lrho0}\
\end{align}

Taking the double scalar product of Eq.~\eqref{eq:FLdeltauab} with ${{\bf e}_{\rm l}}$ and ${\bf e}_{{\rm t}_k}$ gives
\begin{align}
&e_{\rm l}^ae_{{\rm t}_k}^b\left\{\left[ z \delta \tilde{u}^{ab}({\bf q},z)-{\rm i}\frac{\zeta}{2}q^3 \delta \tilde u_\sigma({\bf q},z)G_T^{adef}e_{\rm l}^be_{\rm l}^de_{\rm l}^ee_\sigma^f-{\rm i}\frac{\zeta}{2}q^3 \delta \tilde u_\sigma({\bf q},z)G_T^{bdef}e_{\rm l}^ae_{\rm l}^de_{\rm l}^ee_\sigma^f\right] -q^aq^b{\cal D}_y{\delta} \tilde y({\bf q},z)\right.\notag\\
&\left.+\frac{ \xi^{\prime}}{T}q^aq^b\delta \tilde{T}({\bf q},z)+\frac{{\rm i}q}{2} \delta \tilde v_\sigma({\bf q},z)\left[ e_{\rm l}^ae_\sigma^b+ e_{\rm l}^be_\sigma^a\right]\right\}=e_{\rm l}^ae_{{\rm t}_k}^b\delta {u}^{ab}({\bf q},0)\;.
\end{align}
The terms involving the scalar product ${\bf e}_{\rm l}\cdot {\bf e}_{{\rm t}_k}=0$ vanish exactly. The non-trivial contributions are
\begin{align}
	e_{\rm l}^ae_{{\rm t}_k}^b\delta\tilde u^{ab}({\bf q},z) &=-\frac{{\rm i}}{2}q \delta \tilde u_{{\rm t}_k}({\bf q},z)\;,\\
	e_{\rm l}^ae_{{\rm t}_k}^b\left[-{\rm i}\frac{\zeta}{2}q^3 \delta \tilde u_\sigma({\bf q},z)G_T^{bdef}e_{\rm l}^ae_{\rm l}^de_{\rm l}^ee_\sigma^f\right]&=-{\rm i}\frac{\zeta G_{T,{{\rm t}_k}}}{2}q^3 \delta \tilde u_{{\rm t}_k}({\bf q},z) \;,\\
	e_{\rm l}^ae_{{\rm t}_k}^b\left[\frac{{\rm i}q}{2} \delta \tilde v_\sigma({\bf q},z)e_{\rm l}^ae_\sigma^b+\frac{{\rm i}q}{2} \delta \tilde v_\sigma({\bf q},z)e_{\rm l}^be_\sigma^a\right]&=\frac{{\rm i}q}{2} \delta \tilde v_{{\rm t}_k}({\bf q},z)\;,
\end{align}
giving the transverse equations~\eqref{eq:tu} for $\delta \tilde u^{ab}$. The double scalar product of Eq.~\eqref{eq:FLdeltauab} with ${{\bf e}_{{\rm t}_k}}$ and ${\bf e}_{{\rm t}_j}$ vanishes.

\subsubsection{Linear combination of the longitudinal equations} 

To simplify the right-hand sides of the four longitudinal equations,~\eqref{eq:lv},~\eqref{eq:ly},~\eqref{eq:lT0}, and~\eqref{eq:lrho0}, we take linear combinations. Specifically, summing Eqs.~\eqref{eq:ly} and~\eqref{eq:lrho0} gives Eq.~\eqref{eq:lrho}. Furthermore, Eq.~\eqref{eq:lT} is obtained by summing Eq.~\eqref{eq:lT0} with Eq.~\eqref{eq:ly} multiplied by $\left(\frac{\alpha B_T}{\rho}-\varsigma_y\right)$ and Eq.~\eqref{eq:lrho} multiplied by $\alpha B_T/\rho^2$. These  combinations are required to cast the system in the matrix-vector form of Eq.~\eqref{eq:M-phi=phi}.

\section{Solution of the transverse equations} 
\label{app:transverse}
The transverse equations are solved by matrix inversion to obtain their spectral functions. The dispersion relations,  derived from the poles of these functions, correspond to acoustic waves propagating in the transverse directions. We then derive the propagation velocities and damping coefficients for these modes and compare them to those of a perfect crystal.

The set of transverse equations~\eqref{eq:tv}-\eqref{eq:tu} is cast in a matrix form as
\begin{align}\label{eq:lin-vti-ui}
\left[
\begin{array}{cc}
z+ q^2\frac{\eta_{{\rm t}_k}}{\rho} &  \frac{B_{T,{\rm t}_k}}{\rho}q^2    \\
-1 & z+q^2G_{T,{\rm t}_k} \zeta 
\end{array}
\right]
\left[
\begin{array}{l}
\delta \tilde v_{{\rm t}_k}({\bf q},z) \\
\delta \tilde u_{{\rm t}_k}({\bf q},z)\\
\end{array}
\right]
=
\left[
\begin{array}{l}
\delta v_{{\rm t}_k}({\bf q},0) \\
\delta u_{{\rm t}_k}({\bf q},0)\\
\end{array}
\right].
\end{align}
The solution is obtained by matrix inversion
\begin{align}
\left[
\begin{array}{l}
\delta \tilde v_{{\rm t}_k}({\bf q},z) \\
\delta \tilde u_{{\rm t}_k}({\bf q},z)\\
\end{array}
\right]
&=\frac{1}{\left(z+q^2G_{T,{\rm t}_k} \zeta \right)\left(z+q^2 \frac{\eta_{{\rm t}_k}}{\rho}\right)+ q^2\frac{B_{T,{\rm t}_k}}{\rho}}
\left[
\begin{array}{l}
\left(z+q^2G_{T,{\rm t}_k} \zeta \right)\delta v_{{\rm t}_k}({\bf q},0)- q^2\frac{B_{T,{\rm t}_k}}{\rho}\delta u_{{\rm t}_k}({\bf q},0) \\
\left(z+ q^2\frac{\eta_{{\rm t}_k}}{\rho}\right)\delta u_{{\rm t}_k}({\bf q},0)+\delta v_{{\rm t}_k}({\bf q},0)\\
\end{array}
\right]\;.
\end{align}
With Onsager's hypothesis of regression of fluctuations~\cite{O31b} and the fact that the variables $\delta v_{{\rm t}_k}$ and $\delta u_{{\rm t}_k}$ are statistically independent, we find two pairs of correlation functions
\begin{align}
	\frac{\langle\delta \hat{\tilde v}_{{\rm t}_k}({\bf q},z)\delta \hat v^*_{{\rm t}_k}({\bf q},0) \rangle_{\rm eq}}{\langle\delta \hat v_{{\rm t}_k}({\bf q},0)\delta \hat v^*_{{\rm t}_k}({\bf q},0) \rangle_{\rm eq}}&=\frac{z+ q^2G_{T,{\rm t}_k} \zeta}{\left(z+q^2G_{T,{\rm t}_k} \zeta \right)\left(z+q^2 \frac{\eta_{{\rm t}_k}}{\rho}\right)+q^2 \frac{B_{T,{\rm t}_k}}{\rho}}\,,\label{eq:vvwovvt}\\
	\frac{\langle\delta\hat{\tilde  u}_{{\rm t}_k}({\bf q},z)\delta \hat u^*_{{\rm t}_k}({\bf q},0) \rangle_{\rm eq}}{\langle\delta \hat u_{{\rm t}_k}({\bf q},0)\delta \hat u^*_{{\rm t}_k}({\bf q},0) \rangle_{\rm eq}}&=\frac{z+q^2 \frac{\eta_{{\rm t}_k}}{\rho}}{\left(z+q^2G_{T,{\rm t}_k} \zeta \right)\left(z+ q^2\frac{\eta_{{\rm t}_k}}{\rho}\right)+q^2\frac{B_{T,{\rm t}_k}}{\rho}}\,,
\end{align}
(no summation over ${\rm t}_k$).
The correlation functions of the transverse momentum fluctuations are defined as
\begin{align}
	C_{{\rm t}_k}({\bf q},t) & \equiv \frac{N}{V^2}\left\langle\delta \hat v_{{\rm t}_k}({\bf q},t)\delta\hat v^*_{{\rm t}_k}({\bf q},0) \right\rangle_{\rm eq}\;,
\end{align}
and the corresponding spectral functions as 
\begin{align}
	J_{{\rm t}_k}({\bf q},\omega) & \equiv \frac{N}{V^2}\left\langle\delta \hat{ v}_{{\rm t}_k}({\bf q},\omega)\delta\hat v^*_{{\rm t}_k}({\bf q},0) \right\rangle_{\rm eq}\;.
\end{align}
The spectral functions of the transverse momentum fluctuations are obtained from the relation  $J_{{\rm t}_k}({\bf q},\omega)  =2\,{\rm Re}\left[ \tilde C_{{\rm t}_k}({\bf q},z={\rm i}\omega)\right]$, where $ \tilde C_{{\rm t}_k}({\bf q},z)$ is the Laplace transform of the correlation functions $C_{{\rm t}_k}({\bf q},t)$. We find
\begin{align}
	\frac{J_{{\rm t}_k}({\bf q},\omega)}{C_{{\rm t}_k}({\bf q},0)} & =2 \frac{\eta_{{\rm t}_k}q^2\omega^2/\rho+G_{T,{\rm t}_k} \zeta\left({B_{T,{\rm t}_k}}+G_{T,{\rm t}_k} \zeta{\eta_{{\rm t}_k}}q^2\right)q^4/\rho}{\left(\omega^2-{B_{T,{\rm t}_k}}q^2/\rho-G_{T,{\rm t}_k} \zeta{\eta_{{\rm t}_k}}q^4/\rho\right)^2+\left(\eta_{{\rm t}_k}/\rho+G_{T,{\rm t}_k} \zeta\right)^2q^4\omega^2}\;.\label{eq:JtoCt}
\end{align}
In the absence of vacancies, the  spectral functions of the perfect crystal, found in Ref.~\cite{MG24b}, are recovered.

To characterize the modes, we look for the dispersion relations.  The latter are obtained from the poles of the spectral functions, given by the roots of the determinant of the matrix in Eq.~\eqref{eq:lin-vti-ui} as
\begin{align}
	z_{*} &=\pm {\rm i}\sqrt{\frac{B_{T,{\rm t}_k}}{\rho}}q\left[1-\frac{\rho}{8B_{T,{\rm t}_k}}\left(\frac{\eta_{{\rm t}_k}}{\rho}-G_{T,{\rm t}_k} \zeta \right)^2q^2  +\mathcal{O}(q^4) \right]-\frac{1}{2}\left(\frac{\eta_{{\rm t}_k}}{\rho}+G_{T,{\rm t}_k} \zeta \right)q^2\;.
\end{align}
In the hydrodynamic limit, the dispersion relations are thus given by
\begin{align}
	z_{{\rm t}_k\pm}(q)&=\pm {\rm i} c_{{\rm t}_k} q-\Gamma_{{\rm t}_k}q^2\;,
\end{align}
where $c_{{\rm t}_k\pm}$ are the transverse sound velocities, given in Eq.~\eqref{eq:tcs}, and $\Gamma_{{\rm t}_k}$ the transverse sound damping coefficients, given in Eq.~\eqref{eq:Gammatfull}. We  identify a pair of acoustic modes propagating in each transverse direction ${\rm t}_k$. At leading order in $q$,  the propagation speed is given by the same expression as for a perfect crystal~\cite{MG24b}. The next-order contribution to the velocity, of order $q^2$, includes a term involving the vacancy conductivity $\zeta$. Similarly, the sound damping coefficient $\Gamma_{{\rm t}_k}$ also differs from the expression for the perfect crystal due to a term involving $\zeta$. As expected, in the absence of vacancies the propagation velocity and the damping coefficient of the perfect crystal are recovered, since $\zeta$ is of order ${\cal O}(y)$~\cite{MG25}. 

\section{Solution of the longitudinal equations} 
\label{app:longeqs}

We derive the components $(1,1)$ and $(4,1)$ of the cofactor matrix  $\boldsymbol{\mathsf C}_{M_{\rm l}}(q,z)$, and the determinant of the matrix $\boldsymbol{\mathsf M}_{\rm l}(q,z)$. We also compute the roots of the determinant, giving the poles of the dynamic structure factor.

\subsection{Inversion of the matrix $\boldsymbol{\mathsf M}_{\rm l}(q,z) $}
For the component $(1,1)$ of the cofactor matrix, we find
\begin{align}
	\left[\boldsymbol{\mathsf C}_{M_{\rm l}}(q,z)\right]_{1,1}&= z^3+z^2C^{(1,1)}_{2,2}q^2+z\left(C^{(1,1)}_{1,2}q^2+C^{(1,1)}_{1,4}q^4\right)+C^{(1,1)}_{0,4}q^4+C^{(1,1)}_{0,6}q^6\;,\label{eq:C11}
\end{align}
where
\begin{align}
	C^{(1,1)}_{2,2}&\equiv\frac{ \kappa^{\prime} }{\rho c_v}+ \frac{\eta_{\rm l}}{\rho}+{\cal D}_y+\zeta G_{T,{\rm l}}+\frac{\xi^\prime\left(\rho\varsigma_y-\alpha B_T \right) }{\rho c_v}\;,\\
	C^{(1,1)}_{1,2}&\equiv\frac{B_T   \left({\gamma-1}\right)}{\rho}\;,\\
	C^{(1,1)}_{1,4}&\equiv  \frac{ \kappa^{\prime}\eta_{\rm l} }{\rho^2 c_v}+\frac{ \kappa^{\prime} \left({\cal D}_y+\zeta G_{T,{\rm l}}\right)}{\rho c_v}+\frac{\eta_{\rm l}\left({\cal D}_y+\zeta G_{T,{\rm l}}\right)}{\rho}+\frac{\xi'\eta_{\rm l}\left(\rho\varsigma_y-\alpha B_T \right)}{\rho^2 c_v}-\frac{\xi'\left( {G_{T,{\rm l}} \xi}+ {\cal K}_y\right)}{T\rho c_v} \;,\\
	C^{(1,1)}_{0,4}&\equiv \frac{\left(\gamma-1\right)\left(\pi_y+ B_{T,{\rm l}}\right)\xi'}{\alpha\rho T }+\frac{ \left( {\cal D}_y+\zeta G_{T,{\rm l}}\right)B_T   \left({\gamma-1}\right) }{\rho }\,,\\
	C^{(1,1)}_{0,6}&\equiv \frac{ \kappa^{\prime}\eta_{\rm l}\left({\cal D}_y+\zeta G_{T,{\rm l}}\right) }{\rho^2 c_v}-\frac{\eta_{\rm l}\xi'\left( {G_{T,{\rm l}} \xi}+ {\cal K}_y\right)}{T\rho^2 c_v}\;.
\end{align}

For the component $(4,1)$ of the cofactor matrix, we find
\begin{align}
	\left[\boldsymbol{\mathsf C}_{M_{\rm l}}(q,z)\right]_{4,1}&= z C^{(4,1)}_{1,2}q^2+C^{(4,1)}_{0,4}q^4\;,\label{eq:C41}
\end{align}
where
\begin{align}
	C^{(4,1)}_{1,2}&\equiv-\left({\pi_y+ B_{T,{\rm l}}}\right) \;,\\
	C^{(4,1)}_{0,4}&\equiv-\frac{1}{\rho c_v}\left\{{\alpha B_T}\left[{G_{T,{\rm l}} \xi}+ {\cal K}_y+T\left( {\cal D}_y+\zeta G_{T,{\rm l}}\right) \left(\rho\varsigma_y-\alpha B_T \right)\right] +\left[\kappa^{\prime}+\xi^\prime\left(\rho\varsigma_y-\alpha B_T \right) \right]\left({\pi_y+ B_{T,{\rm l}}}\right) \right\}\;.
\end{align}

\subsection{Determinant of $\boldsymbol{\mathsf M}_{\rm l}(q,z)$} 

We find the determinant of the matrix $\boldsymbol{\mathsf M}_{\rm l}(q,z)$ given in Eq.~\eqref{eq:lin-rho-T-vl-qz} as
\begin{align}
	{\rm det}\ \boldsymbol{\mathsf M}_{\rm l}(q,z)  
	&=z^4 + z^3 C_{3,2}q^2+z^2\left(C_{2,2}q^2+C_{2,4}q^4\right)+z\left(C_{1,4}q^4+C_{1,6}q^6\right)+C_{0,6}q^6\;,\label{eq:detMl}
\end{align}
where
\begin{align}
	C_{3,2}&\equiv  \frac{ \kappa^{\prime} }{\rho c_v}+ \frac{\eta_{\rm l}}{\rho}+{\cal D}_y+\zeta G_{T,{\rm l}}+\frac{\xi^\prime\left(\rho\varsigma_y-\alpha B_T \right) }{\rho c_v}=C^{(1,1)}_{2,2}\;,\\
	C_{2,2}&\equiv\frac{B_{T,{\rm l}} +B_T   \left({\gamma-1}\right)}{\rho}\;,\\
	C_{2,4}&\equiv\frac{ \kappa^{\prime}\eta_{\rm l} }{\rho^2 c_v}+\frac{ \kappa^{\prime} \left({\cal D}_y+\zeta G_{T,{\rm l}}\right)}{\rho c_v}+\frac{\eta_{\rm l}\left({\cal D}_y+\zeta G_{T,{\rm l}}\right)}{\rho}+\frac{\xi'\eta_{\rm l}\left(\rho\varsigma_y-\alpha B_T \right)}{\rho^2 c_v}-\frac{\xi'\left( {G_{T,{\rm l}} \xi}+ {\cal K}_y\right)}{T\rho c_v}=C^{(1,1)}_{1,4}\;,\\
	C_{1,4}&\equiv   {\cal D}_y\frac{{B_{T,{\rm l}}} +B_T   \left({\gamma-1}\right)}{\rho}+ \frac{ \kappa^{\prime}B_{T,{\rm l}}}{\rho^2 c_v}+\frac{\pi_y}{\rho}\left(\frac{\alpha B_T\xi'}{\rho c_v }- \zeta G_{T,{\rm l}} \right)+\frac{{G_{T,{\rm l}} \xi}{\alpha B_T }}{\rho^2c_v}+\varsigma_y\frac{\xi' B_{T,{\rm l}}+\zeta T \alpha B_T G_{T,{\rm l}}}{\rho c_v} \;,\\
	C_{1,6}&\equiv \frac{ \kappa^{\prime}\eta_{\rm l}\left({\cal D}_y+\zeta G_{T,{\rm l}}\right) }{\rho^2 c_v}-\frac{\eta_{\rm l}\xi'\left( {G_{T,{\rm l}} \xi}+ {\cal K}_y\right)}{T\rho^2 c_v}=C^{(1,1)}_{0,6}\;,\\
	C_{0,6}&\equiv\frac{1}{\rho^2 c_v}\left[-{\cal K}_y {B_{T,{\rm l}}}\left(\frac{\xi'}{T}\right)+ {\cal D}_y{G_{T,{\rm l}} \xi}{\alpha B_T }- \kappa^{\prime} \pi_y \zeta G_{T,{\rm l}}+{\cal D}_y{ \kappa^{\prime}}{B_{T,{\rm l}}}-{\zeta G_{T,{\rm l}}}\alpha B_T  {\cal K}_y+{\xi}\pi_y G_{T,{\rm l}}\left(\frac{\xi'}{T}\right) \right]\;.
\end{align}

\subsection{Roots of the determinant and dispersion relations}  
\label{app:poles_DSF}

The roots $z_*$ of the determinant~\eqref{eq:detMl} are obtained by solving ${\rm det}\ \boldsymbol{\mathsf M}_{\rm l}(q,z_*)=0 $ order by order in $q$ using the expansion $z_*=A+Bq+Cq^2+\cdots$ up to second order.

\paragraph{Zero order.} 
The coefficient $A$ is found at zero order in $q$ by the condition
\begin{align}
A^4+{\cal O}(q^2)	&=0\;,
\end{align}
which implies that $A=0$, as expected for slow modes.

\paragraph{First order.} 
The coefficient $B$ is found at first order in $q$, using that $A=0$, by the condition
\begin{align}
B^2\left(B^2+C_{2,2}\right)+\mathcal{O}(q)&=0\;.
\end{align}
We find two solutions with $B=0$ and two solutions with
\begin{align}
	B&=\pm {\rm i} \sqrt{\frac{{B_{T,{\rm l}}} +B_T   \left({\gamma-1}\right)}{\rho}}=\pm {\rm i} c_{\rm l}\;,
\end{align}
where we define the longitudinal velocity  $c_{\rm l}^2\equiv \left[{B_{T,{\rm l}}} +B_T   \left({\gamma-1}\right)\right]/\rho$. This is in agreement with earlier results~\cite{MG24b,MG25}.

The analysis at linear order in $q$ gives two propagating modes with velocity $\pm c_{\rm l}$, corresponding to a pair of longitudinal acoustic modes, and two additional purely diffusive modes.

\paragraph{Second order.}
Starting with the two solutions $B=0$, which correspond to purely diffusive modes, the coefficient $C$ is given by the condition
\begin{align}
	C^2C_{2,2}+CC_{1,4}+C_{0,6}+{\cal O}(q^2)&=0\;, 
\end{align}
which is solved by
\begin{align}
	C_{\pm}&=\frac{-C_{1,4}\pm\sqrt{C^2_{1,4}-4C_{2,2}C_{0,6}}}{2C_{2,2}}\;.\label{eq:Cpm}
\end{align}
To characterize the two diffusive modes, we compute Eq.~\eqref{eq:Cpm} at leading order in the molar fraction of vacancies $y$. Using that ${\cal D}_y , {\cal K}_y , \pi_y \sim\mathcal{O}(y^0)$ and $\xi, \zeta \sim\mathcal{O}(y)$~\cite{MG25}, we find
\begin{align}
	C_{\pm}&=\frac{-\left( {\cal D}_yc_{\rm l}^2+ \frac{ \kappa B_{T,{\rm l}}}{\rho^2 c_v}\right)\pm\left(\frac{ \kappa B_{T,{\rm l}}}{\rho^2 c_v}- {\cal D}_yc_{\rm l}^2\right)}{2c_{\rm l}^2}+\mathcal{O}(y)=\left\{\begin{array}{c}
-{\cal D}_y+\mathcal{O}(y)\\
-\chi_{0}+\mathcal{O}(y)\\
\end{array}\;,
\right.
\end{align}
where $\chi_{0}=\frac{ \kappa B_{T,{\rm l}}}{c_{\rm l}^2\rho^2 c_v}$ is the perfect crystal heat diffusivity~\cite{MG24b} and ${\cal D}_y$ the leading order contribution of the vacancy diffusivity~\cite{MG25}.  We thus identify the two diffusive modes: one mode of heat diffusion with the diffusivity $\chi$ given in Eq.~\eqref{eq:chifull} (which is equal to the perfect crystal value $\chi_{0}$ in the limit of vanishing $y$), and one mode of vacancy diffusion with the diffusivity $D_{\rm vac}$, given in Eq.~\eqref{eq:Dvacfull} (which is given by the vacancy diffusion coefficient ${\cal D}_y$ at leading order in $y$, in agreement with Ref.~\cite{MG25}).

Considering the solutions $B=\pm {\rm i}c_{\rm l}$, that correspond to the acoustic modes propagating in the longitudinal direction with velocity $c_{\rm l}$, the coefficient $C$ is given by the condition
\begin{align}
	\left(c_{\rm l}^4-c_{\rm l}^2C_{2,2}\right)q^4+\left(\mp 4{\rm i} c_{\rm l}^3C\mp {\rm i} c_{\rm l}^3C_{3,2}\pm 2{\rm i}c_{\rm l}CC_{2,2}\pm {\rm i}c_{\rm l}C_{1,4}\right)q^5+\mathcal{O}(q^6)&=0\;.
\end{align}
The term of order $q^4$ vanishes exactly since $C_{2,2}=c_{\rm l}^2$. From the order $q^5$, the coefficient $C$ is given by 
\begin{align}
	C&=\frac{1}{2}\left(\frac{C_{1,4}}{c_{\rm l}^2}-C_{3,2}\right)\notag\\
		&=- \frac{\eta_{\rm l}}{2\rho}-\frac{ \kappa^{\prime} }{2\rho c_v}\left(1-\frac{ B_{T,{\rm l}}}{\rho c_{\rm l}^2}\right)+\frac{ G_{T,{\rm l}} \xi \alpha B_T }{2\rho^2c_vc_{\rm l}^2}-\frac{1}{2}\left(\zeta G_{T,{\rm l}}-\frac{\xi^\prime\alpha B_T}{\rho c_v}\right)\left(1+\frac{\pi_y}{\rho c_{\rm l}^2}-\frac{\varsigma_yT\alpha B_T}{\rho c_v c_{\rm l}^2}\right)\;.\label{eq:Ccoeff}
\end{align}
At leading order in $y$, we have
\begin{align}
 C&= -\frac{1}{2}\left[ \frac{\eta_{\rm l}}{\rho}+\frac{ \kappa }{\rho c_p}\frac{\gamma}{1+  \left({\gamma-1}\right)^{-1}\frac{B_{T,{\rm l}}}{B_T}}\right]+\mathcal{O}(y)=-\Gamma_{\rm l 0}+\mathcal{O}(y)\;,
\end{align}
where  $\Gamma_{\rm l 0}$ is the perfect crystal longitudinal acoustic damping coefficient~\cite{MG24b}.  The last two modes are thus a pair of acoustic waves propagating in the longitudinal direction with velocity $c_{\rm l}$ and a damping coefficient $\Gamma_{\rm l}\equiv -C$, given in Eq.~\eqref{eq:Gammalfull}. This agrees with the result of Ref.~\cite{MG25} and equals the perfect crystal value $\Gamma_{\rm l 0}$ in the limit of vanishing $y$.

\subsection{Small $q$ expansion of the spectrum}
\label{app:resonances}

The inverse Laplace transform of Eq.~\eqref{eq:Sks} gives the intermediate scattering function $F({\bf q},t)$ as
\begin{align}
	F({\bf q},t)  &=\frac{1}{2\pi{\rm i}}\int_{c-{\rm i}\infty}^{c+{\rm i}\infty}\text{d}z\frac{\left[\boldsymbol{\mathsf C}_{M_{\rm l}}(q,z)\right]_{1,1}{\rm e}^{zt}}{\prod_{i={\rm H},{\rm V},{\rm l}\pm}(z-z_i)}S({\bf q})+\frac{1}{2\pi{\rm i}}\int_{c-{\rm i}\infty}^{c+{\rm i}\infty}\text{d}z\frac{\left[\boldsymbol{\mathsf C}_{M_{\rm l}}(q,z)\right]_{4,1}{\rm e}^{zt}}{\prod_{i={\rm H},{\rm V},{\rm l}\pm}(z-z_i)}S_{\rm nv}({\bf q})\notag\\
	&=S({\bf q})\sum_{i={\rm H},{\rm V},{\rm l}\pm}\lim_{z\rightarrow z_i}\frac{\left[\boldsymbol{\mathsf C}_{M_{\rm l}}(q,z)\right]_{1,1}(z-z_i)}{\prod_{j={\rm H},{\rm V},{\rm l}\pm}(z-z_j)}{\rm e}^{z|t|}+S_{\rm nv}({\bf q})\sum_{i={\rm H},{\rm V},{\rm l}\pm}\lim_{z\rightarrow z_i}\frac{\left[\boldsymbol{\mathsf C}_{M_{\rm l}}(q,z)\right]_{4,1}(z-z_i)}{\prod_{j={\rm H},{\rm V},{\rm l}\pm}(z-z_j)}{\rm e}^{z|t|}\;.\label{eq:DSFstILT}
\end{align}
We compute each contribution in the sums at leading order in the wave number and in the molar fraction of vacancies.

\paragraph{Vacancy mode.} We consider first the vacancy diffusion term $z_{\rm V}$. We have
\begin{align}
\left[\boldsymbol{\mathsf C}_{M_{\rm l}}(q,z_{\rm V})\right]_{1,1}&=(-{D}_{\rm vac}q^2)^3+(-{D}_{\rm vac}q^2)^2C^{(1,1)}_{2,2}q^2+(- {D}_{\rm vac}q^2)\left(C^{(1,1)}_{1,2}q^2+C^{(1,1)}_{1,4}q^4\right)+C^{(1,1)}_{0,4}q^4+C^{(1,1)}_{0,6}q^6\notag\\
&=\frac{\gamma-1}{\rho}\left[\frac{\left(\pi_y+ B_{T,{\rm l}}\right)\xi'}{\alpha T }+ (\zeta G_{T,{\rm l}}-\Delta {D}_{\rm vac})B_T \right]q^4 +\mathcal{O}(q^6)\;,\\
\left[\boldsymbol{\mathsf C}_{M_{\rm l}}(q,z_{\rm V})\right]_{4,1}&=(-{\cal D}_yq^2)C^{(4,1)}_{1,2}q^2+C^{(4,1)}_{0,4}q^4\notag\\
&=-\frac{\gamma-1}{\rho}\left\{\frac{\rho}{T\alpha}\left[{\cal K}_y+T {\cal D}_y\left(\rho\varsigma_y-\alpha B_T \right)\right]+\frac{\rho}{T\alpha^2B_T}\left({\kappa}-{\rho c_v }{\cal D}_y \right)\left({\pi_y+ B_{T,{\rm l}}}\right)+{\cal O}(y) \right\}q^4\;,
\end{align}
where $\Delta {D}_{\rm vac}\equiv {D}_{\rm vac}-{\cal D}_y $. Moreover
\begin{align}
	\lim_{z\rightarrow z_{\rm V}}\frac{(z-z_{\rm V})}{\prod_{j={\rm H},{\rm V},{\rm l}\pm}(z-z_j)}&=(\chi-{D}_{\rm vac})^{-1}c_{\rm l}^{-2}q^{-4}+\mathcal{O}(q^{-2})\;,\label{eq:den_zv}
\end{align}
and, therefore, the contributions from $z_{\rm V}$ read
\begin{align}
	\lim_{z\rightarrow z_{\rm V}}\frac{\left[\boldsymbol{\mathsf C}_{M_{\rm l}}(q,z)\right]_{1,1}(z-z_{\rm V})}{\prod_{j={\rm H},{\rm V},{\rm l}\pm}(z-z_j)}{\rm e}^{z|t|}&=\frac{\gamma-1}{\rho c_{\rm l}^2}\left[\frac{\left(\pi_y+ B_{T,{\rm l}}\right)\xi'}{\alpha T(\chi-{D}_{\rm vac}) }+\frac{(\zeta G_{T,{\rm l}}-\Delta {D}_{\rm vac})B_T}{\chi-{D}_{\rm vac}}+\mathcal{O}(q^2)\right]{\rm e}^{-{D}_{\rm vac}q^2|t|}\;,\\
		\lim_{z\rightarrow z_{\rm V}}\frac{\left[\boldsymbol{\mathsf C}_{M_{\rm l}}(q,z)\right]_{4,1}(z-z_{\rm V})}{\prod_{j={\rm H},{\rm V},{\rm l}\pm}(z-z_j)}{\rm e}^{z|t|}&=-\frac{\gamma-1}{\rho c_{\rm l}^2}\left[\rho\frac{{\cal K}_y+T {\cal D}_y\left(\rho\varsigma_y-\alpha B_T \right)}{T\alpha(\chi_0-{\cal D}_y)}+\frac{\rho \left({\kappa}-{\rho c_v }{\cal D}_y \right)\left({\pi_y+ B_{T,{\rm l}}}\right)}{T\alpha^2B_T(\chi_0-{\cal D}_y)}+ \mathcal{O}(y)\right]{\rm e}^{-D_{\rm vac}q^2|t|}\;.
\end{align}

\paragraph{Heat mode.} For the heat diffusion term $z_{\rm H}$, we have
\begin{align}
\left[\boldsymbol{\mathsf C}_{M_{\rm l}}(q,z_{\rm H})\right]_{1,1}&=(-\chi q^2)^3+(-\chi q^2)^2C^{(1,1)}_{2,2}q^2+(-\chi q^2)\left(C^{(1,1)}_{1,2}q^2+C^{(1,1)}_{1,4}q^4\right)+C^{(1,1)}_{0,4}q^4+C^{(1,1)}_{0,6}q^6\notag\\
&=-\frac{(\chi_0-{\cal D}_y)B_T   \left({\gamma-1}\right)}{\rho}q^4+\mathcal{O}(yq^4)+\mathcal{O}(q^6)\;,\\
\left[\boldsymbol{\mathsf C}_{M_{\rm l}}(q,z_{\rm H})\right]_{4,1}&=(-\chi_{0} q^2)C^{(4,1)}_{1,2}q^2+C^{(4,1)}_{0,4}q^4\notag\\
&=-(\gamma-1)\left[\frac{{\cal K}_y+T {\cal D}_y\left(\rho\varsigma_y-\alpha B_T \right)}{T\alpha}+\frac{\left(\kappa-\rho c_v \chi_{0} \right)\left({\pi_y+ B_{T,{\rm l}}}\right) }{T\alpha^2B_T}+{\cal O}(y)\right]q^4\;,
\end{align}
and
\begin{align}
	\lim_{z\rightarrow z_{\rm H}}\frac{(z-z_{\rm H})}{\prod_{j={\rm H},{\rm V},{\rm l}\pm}(z-z_j)}&=-(\chi_0-{\cal D}_y)^{-1}c_{\rm l}^{-2}q^{-4}+\mathcal{O}(yq^{-4})+\mathcal{O}(q^{-2})\;,
	\label{eq:den_zh}
\end{align}
and, therefore, the contributions from $z_{\rm H}$, at leading order in $y$ read
\begin{align}
	\lim_{z\rightarrow z_{\rm H}}\frac{\left[\boldsymbol{\mathsf C}_{M_{\rm l}}(q,z)\right]_{1,1}(z-z_{\rm H})}{\prod_{j={\rm H},{\rm V},{\rm l}\pm}(z-z_j)}{\rm e}^{z|t|}
	&=\frac{\gamma-1}{\rho c_{\rm l}^2}\left[B_T+\mathcal{O}(y)+\mathcal{O}(q^2)\right]{\rm e}^{-\chi q^2|t|}\;,\label{eq:ISF_zH}\\
\lim_{z\rightarrow z_{\rm H}}\frac{\left[\boldsymbol{\mathsf C}_{M_{\rm l}}(q,z)\right]_{1,1}(z-z_{\rm H})}{\prod_{j={\rm H},{\rm V},{\rm l}\pm}(z-z_j)}{\rm e}^{z|t|}&=\frac{\gamma-1}{c_{\rm l}^2}\left[\frac{{\cal K}_y+T {\cal D}_y\left(\rho\varsigma_y-\alpha B_T \right)}{T\alpha(\chi_{0}-{\cal D}_y)}+\frac{\left(\kappa-\rho c_v \chi_{0} \right)\left({\pi_y+ B_{T,{\rm l}}}\right)}{T\alpha^2B_T(\chi_{0}-{\cal D}_y)}+\mathcal{O}(y)+\mathcal{O}(q^2)\right]{\rm e}^{-\chi q^2|t|}\;.
\end{align}
Remarkably, the coefficients $(\chi_0-{\cal D}_y)$ cancel exactly in Eq.~\eqref{eq:ISF_zH}.

\paragraph{Longitudinal acoustic modes.} For the propagating sound mode $z_{\rm l+}$, we have
\begin{align}
\left[\boldsymbol{\mathsf C}_{M_{\rm l}}(q,z_{{\rm l}+})\right]_{1,1}&=(-{\rm i}c_{\rm l}^3q^3+3c_{\rm l}^2\Gamma_{\rm l} q^4+3{\rm i}c_{\rm l}\Gamma_{\rm l}^2q^5-\Gamma_{\rm l}^3q^6)+(-c_{\rm l}^2q^2-2{\rm i}c_{\rm l}\Gamma_{\rm l} q^3+\Gamma_{\rm l}^2q^4)C^{(1,1)}_{2,2}q^2\notag\\
&+({\rm i}c_{\rm l}q-\Gamma_{\rm l} q^2)\left(C^{(1,1)}_{1,2}q^2+C^{(1,1)}_{1,4}q^4\right)+C^{(1,1)}_{0,4}q^4+C^{(1,1)}_{0,6}q^6\notag\\
&=\left(-{\rm i}c_{\rm l}^3+{\rm i}c_{\rm l}C^{(1,1)}_{1,2}\right)q^3+\left(3c_{\rm l}^2 \Gamma_{\rm l}-c_{\rm l}^2 C^{(1,1)}_{2,2}-\Gamma_{\rm l} C^{(1,1)}_{1,2}+C^{(1,1)}_{0,4} \right)q^4+{\cal O}(q^5)\;,
\end{align}
where we compute
\begin{align}
	-{\rm i}c_{\rm l}^3+{\rm i}c_{\rm l}C^{(1,1)}_{1,2}&=-{\rm i}c_{\rm l}\frac{B_{T,{\rm l}}}{\rho}\;,
\end{align}
and
\begin{align}
	3c_{\rm l}^2 \Gamma_{\rm l }-c_{\rm l}^2 C^{(1,1)}_{2,2}-\Gamma_{\rm l } C^{(1,1)}_{1,2}+C^{(1,1)}_{0,4} &=-\chi_{0} c_{\rm l}^2+\left(\Gamma_{\rm l 0}-{\cal D}_y\right)\frac{B_{T,{\rm l}}}{\rho}+\mathcal{O}(y)\;,
\end{align}
using that	$\chi_{0}-\gamma D_T=D_v-2\Gamma_{\rm l 0}$,  where $ D_T\equiv \kappa/(\rho c_p)$ is the thermal diffusivity  and $D_v\equiv\eta_{\rm l}/\rho$ the longitudinal viscosity. We thus have
\begin{align}
\left[\boldsymbol{\mathsf C}_{M_{\rm l}}(q,z_{{\rm l}+})\right]_{1,1}&=-{\rm i}c_{\rm l}\frac{B_{T,{\rm l}}}{\rho}q^3+\left[-c_{\rm l}^2\chi_{0} +\left(\Gamma_{\rm l 0}-{\cal D}_y\right)\frac{B_{T,{\rm l}}}{\rho} \right]q^4+\mathcal{O}(yq^4)+\mathcal{O}(q^5)\;.
\end{align}
The contribution from $\left[\boldsymbol{\mathsf C}_{M_{\rm l}}(q,z_{{\rm l}+})\right]_{4,1}$ is
\begin{align}
\left[\boldsymbol{\mathsf C}_{M_{\rm l}}(q,z_{{\rm l}+})\right]_{4,1}&=({\rm i}c_{\rm l}q-\Gamma_{\rm l 0} q^2)C^{(4,1)}_{1,2}q^2+C^{(4,1)}_{0,4}q^4\notag\\
&=-{\rm i}c_{\rm l}\left({\pi_y+ B_{T,{\rm l}}}\right)q^3-\left\{\frac{\alpha B_T}{\rho c_v}\left[{\cal K}_y+T {\cal D}_y\left(\rho\varsigma_y-\alpha B_T \right)\right]+\left(\frac{\kappa}{\rho c_v}- \Gamma_{\rm l 0} \right)\left({\pi_y+ B_{T,{\rm l}}}\right)  \right\}q^4+{\cal O}(yq^4)\;.
\end{align}
Moreover, we obtain
\begin{align}
	\lim_{z\rightarrow z_{{\rm l}+}}\frac{(z-z_{{\rm l}+})}{\prod_{j={\rm H},{\rm V},{\rm l}\pm}(z-z_j)}&=({\rm i}c_{\rm l}-\Gamma_{\rm l } q^2+{\rm i}c_{\rm l}+\Gamma_{\rm l } q^2)^{-1}({\rm i}c_{\rm l}-\Gamma_{\rm l } q^2+\chi q^2)^{-1}({\rm i}c_{\rm l}-\Gamma_{\rm l } q^2+{D}_{\rm vac}q^2)^{-1}\notag\\
	&=\left\{2{\rm i}c_{\rm l}^2q^3\left[-c_{\rm l}+{\rm i}q(\chi_{0}+{\cal D}_y-2\Gamma_{\rm l 0})+\mathcal{O}(yq)+\mathcal{O}(q^2)\right]\right\}^{-1}\;,\label{eq:denzl}
\end{align}
and, therefore, the contributions from $z_{\rm l +}$ read
\begin{align}
	\lim_{z\rightarrow z_{\rm l +}}&\frac{\left[\boldsymbol{\mathsf C}_{M_{\rm l}}(q,z)\right]_{1,1}(z-z_{{\rm l}+})}{\prod_{j={\rm H},{\rm V},{\rm l}\pm}(z-z_j)}{\rm e}^{z|t|}=\frac{-{\rm i}c_{\rm l}\frac{B_{T,{\rm l}}}{\rho}q^3\left[1-{\rm i}\frac{ c_{\rm l} \rho q}{B_{T,{\rm l}}}\chi_{0}+\frac{{\rm i}q}{c_{\rm l}}\left(\Gamma_{\rm l 0}-{\cal D}_y\right)+\mathcal{O}(yq)+\mathcal{O}(q^2) \right]}{-2{\rm i}c_{\rm l}^3q^3\left[1-\frac{{\rm i}q}{c_{\rm l}}(\chi_{0}+{\cal D}_y-2\Gamma_{\rm l 0})+\mathcal{O}(yq)+\mathcal{O}(q^2)\right]}{\rm e}^{({\rm i}c_{\rm l}q-\Gamma_{\rm l } q^2)|t|}\notag\\
	&=\frac{B_{T,{\rm l}}}{2\rho c_{\rm l}^2}\left[1+\frac{{\rm i}q}{c_{\rm l}}(D_v-3\Gamma_{\rm l 0})+\mathcal{O}(yq)+\mathcal{O}(q^2)\right]{\rm e}^{({\rm i}c_{\rm l}q-\Gamma_{\rm l } q^2)|t|}\;,\label{eq:ISF_z+}\\
	\lim_{z\rightarrow z_{\rm l +}}&\frac{\left[\boldsymbol{\mathsf C}_{M_{\rm l}}(q,z)\right]_{4,1}(z-z_{{\rm l}+})}{\prod_{j={\rm H},{\rm V},{\rm l}\pm}(z-z_j)}{\rm e}^{z|t|}\notag\\
	&=\frac{-{\rm i}c_{\rm l}\left({\pi_y+ B_{T,{\rm l}}}\right)q^3\left[1-\frac{{\rm i}q}{c_{\rm l}}\frac{\alpha B_T\left[{\cal K}_y+T {\cal D}_y\left(\rho\varsigma_y-\alpha B_T \right)\right]}{\rho c_v\left({\pi_y+ B_{T,{\rm l}}}\right)}-\frac{{\rm i}q}{c_{\rm l}}\left(\frac{\kappa}{\rho c_v}-\Gamma_{\rm l 0} \right)+\mathcal{O}(yq)+\mathcal{O}(q^2) \right]}{-2{\rm i}c_{\rm l}^3q^3\left[1-\frac{{\rm i}q}{c_{\rm l}}(\chi_{0}+{\cal D}_y-2\Gamma_{\rm l 0})+\mathcal{O}(yq)+\mathcal{O}(q^2)\right]}{\rm e}^{({\rm i}c_{\rm l}q-\Gamma_{\rm l } q^2)|t|}\notag\\
&=\frac{\pi_y+ B_{T,{\rm l}}}{2 c_{\rm l}^2}\left\{1-\frac{{\rm i}q}{c_{\rm l}}\frac{\alpha B_T\left[{\cal K}_y+T {\cal D}_y\left(\rho\varsigma_y-\alpha B_T \right)\right]}{\rho c_v\left({\pi_y+ B_{T,{\rm l}}}\right)}+\frac{{\rm i}q}{c_{\rm l}}(D_v+{\cal D}_y-3\Gamma_{\rm l 0})+\mathcal{O}(yq)+\mathcal{O}(q^2)\right\}{\rm e}^{({\rm i}c_{\rm l}q-\Gamma_{\rm l} q^2)|t|}\;,\label{eq:ISF_44z+}
\end{align}
using that $\chi_{0} = \gamma D_T B_{T,{\rm l}}/(\rho c_{\rm l}^2)$. Remarkably, the coefficients ${\cal D}_y$ cancel exactly in Eq.~\eqref{eq:ISF_z+} .

We directly obtain the contributions from $z_{\rm l -}$ by replacing $c_{\rm l}$ with $-c_{\rm l}$  in Eqs.~\eqref{eq:ISF_z+} and~\eqref{eq:ISF_44z+}:
\begin{align}
	\lim_{z\rightarrow z_{\rm l -}}&\frac{\left[\boldsymbol{\mathsf C}_{M_{\rm l}}(q,z)\right]_{1,1}(z-z_{{\rm l}-})}{\prod_{j={\rm H},{\rm V},{\rm l}\pm}(z-z_j)}{\rm e}^{z|t|}=\frac{B_{T,{\rm l}}}{2\rho c_{\rm l}^2}\left[1-\frac{{\rm i}q}{c_{\rm l}}(D_v-3\Gamma_{\rm l 0})+\mathcal{O}(yq)+\mathcal{O}(q^2)\right]{\rm e}^{(-{\rm i}c_{\rm l}q-\Gamma_{\rm l } q^2)|t|}\;,\label{eq:ISF_z-}\\
		\lim_{z\rightarrow z_{\rm l -}}&\frac{\left[\boldsymbol{\mathsf C}_{M_{\rm l}}(q,z)\right]_{4,1}(z-z_{{\rm l}-})}{\prod_{j={\rm H},{\rm V},{\rm l}\pm}(z-z_j)}{\rm e}^{z|t|}\notag\\
	&=\frac{\pi_y+ B_{T,{\rm l}}}{2 c_{\rm l}^2}\left\{1+\frac{{\rm i}q}{c_{\rm l}}\frac{\alpha B_T\left[{\cal K}_y+T {\cal D}_y\left(\rho\varsigma_y-\alpha B_T \right)\right]}{\rho c_v\left({\pi_y+ B_{T,{\rm l}}}\right) }-\frac{{\rm i}q}{c_{\rm l}}(D_v+{\cal D}_y-3\Gamma_{\rm l 0})+\mathcal{O}(yq)+\mathcal{O}(q^2)\right\} {\rm e}^{(-{\rm i}c_{\rm l}q-\Gamma_{\rm l } q^2)|t|}\;. 
\end{align}

\paragraph{Intermediate scattering function.} Since we work in the limit  $0\leq y\ll q_*<1$, and using that $S_{\rm nv}\sim {\cal O} (y)$, we obtain  the intermediate scattering function with Eq.~\eqref{eq:DSFstILT} as
\begin{align}
	\frac{F({\bf q},t)}{S({\bf q})}
	&=\frac{\gamma-1}{\rho c_{\rm l}^2}\left[D_{\rm l}(y)+ \mathcal{O}(y^2)+\mathcal{O}(q^2)\right]{\rm e}^{-{D}_{\rm vac}q^2|t|}+\frac{\gamma-1}{\rho c_{\rm l}^2}\left[B_T+\mathcal{O}(y)+\mathcal{O}(q^2)\right]{\rm e}^{-\chi q^2|t|}\notag\\
	&+\frac{B_{T,{\rm l}}}{2\rho c_{\rm l}^2}\left[1+{\rm i}\frac{q}{c_{\rm l}}(D_v-3\Gamma_{\rm l 0})+\mathcal{O}(y)+\mathcal{O}(q^2)\right]{\rm e}^{({\rm i}c_{\rm l}q-\Gamma_{\rm l } q^2)|t|}\notag\\
	&+\frac{B_{T,{\rm l}}}{2\rho c_{\rm l}^2}\left[1-{\rm i}\frac{q}{c_{\rm l}}(D_v-3\Gamma_{\rm l 0})+\mathcal{O}(y)+\mathcal{O}(q^2)\right]{\rm e}^{(-{\rm i}c_{\rm l}q-\Gamma_{\rm l } q^2)|t|}\notag\\
	&=\frac{1}{1+(\gamma-1)\frac{B_T}{B_{T,{\rm l}}}}\left\{\frac{\gamma-1}{B_{T,{\rm l}}}\left[D_{\rm l}(y)+ \mathcal{O}(y^2)+\mathcal{O}(q^2)\right]{\rm e}^{-{D}_{\rm vac}q^2|t|}\right.+\frac{\gamma-1}{B_{T,{\rm l}}}\left[B_T+\mathcal{O}(y)+\mathcal{O}(q^2)\right]{\rm e}^{-\chi q^2|t|}\notag\\
	&+\left.\left[\cos(qc_{\rm l}|t|)+\frac{3\Gamma_{\rm l 0}-D_v}{c_{\rm l}}q\sin(qc_{\rm l}|t|)+\mathcal{O}(y)+\mathcal{O}(q^2)\right]{\rm e}^{-\Gamma_{\rm l } q^2|t|}\right\}\;,\label{eq:ISF_app}
\end{align}
where the coefficient in front of the vacancy diffusion damping mode $D_{\rm l}(y)$ is given by Eq.~\eqref{eq:coeffDofy}.

By definition, the intermediate scattering function at $t=0$ equals the static structure factor, and the left-hand side of Eq.~\eqref{eq:ISF_app} is  equal to one. The contributions at zero order in $y$ from the heat diffusion mode and the acoustic modes on the right-hand side sums up to one, as shown for the perfect crystal in Ref.~\cite{MG24b}. For $t=0$, the $\mathcal{O}(y)$ corrections of the heat diffusion and the acoustic modes cancel exactly the contribution from the vacancy diffusion mode to have $F({\bf q},t)=S({\bf q})$, as seen from the Fourier transform of the full expression~\eqref{eq:DSF_Full} in Fig.~\ref{Fig:DSF_Theo}. The $\mathcal{O}(y)$ corrections of the acoustic and heat modes  in Eq.~\eqref{eq:ISF_app} are thus relevant in this limit.

%%%%%%%%%%%%%%%%%%%%%%%%%%%%%%%%%%%%%%%%%%%%%%%%%
\section{Static correlation functions in the limit ${\bf q}\to 0$}
\label{app:static-fns}
%%%%%%%%%%%%%%%%%%%%%%%%%%%%%%%%%%%%%%%%%%%%%%%%%

Since Fourier transforms are defined as $x({\bf q})\equiv \int_V \text{d}{\bf r} \, {\rm e}^{{\rm i}{\bf q}\cdot{\bf r}} \, x({\bf r})$, we have that $\lim_{{\bf q}\to 0} x({\bf q})= \int_V  \text{d}{\bf r}  \, x({\bf r})=X$, where $X$ is the extensive quantity corresponding to the density $x$.  Therefore, in the zero wave-vector limit, the static structure factor and the static correlation function between the vacancy molar fraction and the particle density fluctuations are given by
\be
\lim_{{\bf q}\to 0} S({\bf q})=\lim_{{\bf q}\to 0} \frac{1}{Nm^2}\langle\delta \hat\rho({\bf q},0)\delta\hat \rho^*({\bf q},0)\rangle_{\rm eq}=\frac{1}{N}\langle ( \hat N - \langle\hat N\rangle)^2 \rangle_{\rm eq}
\ee
and
\be\label{S_nv(0)}
\lim_{{\bf q}\to 0} S_{\rm nv}({\bf q})=\lim_{{\bf q}\to 0} \frac{1}{Nm^2}\langle\delta \hat y({\bf q},0)\delta \hat \rho^*({\bf q},0)\rangle_{\rm eq}
=\frac{V}{mNN_0}\langle ( \hat N_{\rm v} - \langle\hat N_{\rm v}\rangle)( \hat N - \langle\hat N\rangle) \rangle_{\rm eq}
\ee
in terms of the variance of the fluctuating particle number $\hat N$ and its covariance with the fluctuating vacancy number $\hat N_{\rm v}$, which should be calculated in the grand canonical ensemble at fixed values of the temperature $T$, the volume $V$, the chemical potential $\mu_{\rm a}$ of the particles (i.e., the atoms), and the one $\mu_{\rm v}$ of the vacancies, where $N=\langle\hat N\rangle$ is the mean particle number.  The corresponding thermodynamic grand potential is defined by $J\equiv F-\mu_{\rm a}N-\mu_{\rm v}N_{\rm v}$, where $F$ is the Helmholtz free energy considered in Appendix~A of~Ref.~\cite{MG25}.  The methods of equilibrium statistical mechanics show that the variance and the covariance can be calculated as
\be
\langle ( \hat N - \langle\hat N\rangle)^2 \rangle_{\rm eq} = k_{\rm B}T \, \left(\frac{\partial\langle\hat N\rangle}{\partial\mu_{\rm a}}\right)_{T,V,\mu_{\rm v}}
\ee
and
\be\label{cov(N_v,N)}
\langle ( \hat N_{\rm v} - \langle\hat N_{\rm v}\rangle)( \hat N - \langle\hat N\rangle) \rangle_{\rm eq} = k_{\rm B}T \, \left(\frac{\partial\langle\hat N\rangle}{\partial\mu_{\rm v}}\right)_{T,V,\mu_{\rm a}} ,
\ee
where $\langle\hat N\rangle$ is the equilibrium mean value of the number of atoms in the crystal at fixed values of $(T,V,\mu_{\rm a},\mu_{\rm v})$.  Since the derivative of the mean particle number with respect to the particle chemical potential is related to the isothermal compressibility $\chi_T$ according to $\left(\frac{\partial\langle\hat N\rangle}{\partial\mu_{\rm a}}\right)_{T,V,\mu_{\rm v}}
=n \chi_T \langle \hat N \rangle$,  we recover the following well-known result \cite{BP76,BY80},
\be
\lim_{{\bf q}\to 0} S({\bf q})= n \, k_{\rm B}T \, \chi_T \, .
\ee

Next, the covariance can be calculated using the following formula for the chemical potential of vacancies in the limit $y\to 0$,
\be
\mu_{\rm v} = g_{\rm v} + k_{\rm B}T \, \ln y + \mathcal{O}(y) \, ,
\ee
where $g_{\rm v}(T,v)$ is the intensive Gibbs free energy of formation of one vacancy in the crystal and $v\equiv V/N_0$ is the volume of a lattice site in the crystal \cite{MG25}.  Under the same conditions, the particle chemical potential is given by $\mu_{\rm a} = f_0 - v \left(\frac{\partial f_0}{\partial v}\right)_T  + \mathcal{O}(y)$, where $f_0(T,v)$ is the intensive Helmholtz free energy per atom in the perfect crystal.
Therefore, fixing the temperature $T$, the volume $V$, and the particle chemical potential $\mu_{\rm a}$ also fixes the number $N_0$ of lattice sites in the limit $y\to 0$, where the correction of $\mathcal{O}(y)$ can be neglected.  Accordingly, we have $N=N_0-N_{\rm v}=N_0-N_0 y$, so that $\delta N =- N_0 \delta y$.  Consequently, we find
\be
\left(\frac{\partial\langle\hat N\rangle}{\partial\mu_{\rm v}}\right)_{T,V,\mu_{\rm a}} 
= \left(-\frac{1}{N_0}\frac{\partial\mu_{\rm v}}{\partial y}\right)^{-1} \simeq -\frac{N_0y}{k_{\rm B}T}
\ee
in the limit $y\to 0$. Substituting this result into Eq.~\eqref{cov(N_v,N)}, the zero wave-vector limit of the static correlation function~\eqref{S_nv(0)} becomes
\be\label{S_nv-theory}
\lim_{{\bf q}\to 0} S_{\rm nv}({\bf q})=\frac{V}{mNN_0}\, k_{\rm B}T \, \left(\frac{\partial\langle\hat N\rangle}{\partial\mu_{\rm v}}\right)_{T,V,\mu_{\rm a}} \simeq - \frac{y}{mn}
\ee
at low vacancy density.  Accordingly, this static correlation function is indeed of order $\mathcal{O}(y)$.

For the hard-sphere crystal with one vacancy described in Sec.~\ref{sec:MDHS}, the molecular dynamics methods of Appendix~\ref{app:MDHS} give the values of $\lim_{{\bf q}\to 0}S_{\rm nv}({\bf q})$ that are reported in Table~\ref{Tab:S_nv-HS}.  These values are in good agreement with the theoretical expectation given by Eq.~\eqref{S_nv-theory}, confirming the linear behavior of $S_{\rm nv}$ in $y$ for $y\ll 1$.
For a hard-sphere crystal with $N_0=500$ and $N_{\rm v}=1$, the static structure factor is numerically evaluated to be $\lim_{{\bf q}\to 0}S({\bf q})=(1.63\pm0.08)\times 10^{-2}$, which is close to the corresponding value $\lim_{{\bf q}\to 0}S_{\rm pc}({\bf q})=(1.46\pm0.07)\times 10^{-2}$ for the perfect crystal.

%%%%%%%%%%%%%%%%%%%%%%%%%%%%%%%%%%%%%%%%%%%%%%%%%%%
\begin{table}[t!]
\begin{tabular}{c @{\hskip 1cm} c @{\hskip 1cm} c}
\hline\hline
$N_0$ &   $\lim_{{\bf q}\to 0}S_{\rm nv}({\bf q})$     &    $-y/(mn)$   	 \\
\hline  
$500$ & $(-1.81\pm0.02)\times 10^{-3}$ & $-1.90\times 10^{-3}$ \\              
$864$ & $(-1.00\pm0.07)\times 10^{-3}$ & $-1.10\times 10^{-3}$ \\              
$1372$ &  $(-0.68\pm0.06)\times 10^{-3}$ & $-0.69\times 10^{-3}$ \\              
$2048$ & $(-0.41\pm0.03)\times 10^{-3}$ & $-0.46\times 10^{-3}$ \\              
\hline\hline
\end{tabular}
\caption{Numerical computation of the static correlation function~\eqref{S_nv(0)} between the vacancy molar fraction and the particle density fluctuations in the limit ${\bf q}\to 0$ and compared with the theoretical expectation~\eqref{S_nv-theory} for $y\ll 1$, using the molecular dynamics simulation of hard-sphere systems with $N_0$ lattice sites and $N_{\rm v}=1$ vacancy.  The numerical values are obtained with $N_{\rm stat}=10^4$ for $N_0=500$ and $N_0=864$, $N_{\rm stat}=5\times10^3$ for $N_0=1372$, $N_{\rm stat}=2\times10^3$ for $N_0=2048$, $\Delta t=0.1$, and $n_{\rm step}=100$.  The limit ${\bf q}\to 0$ is taken by linear regression using the dependence in $q^2$ and the two smallest non-vanishing values of $q$.}
\label{Tab:S_nv-HS}
\end{table}
%%%%%%%%%%%%%%%%%%%%%%%%%%%%%%%%%%%%%%%%%%%%%%%%%

%%%%%%%%%%%%%%%%%%%%%%%%%%%%%%%%%%%%%%%%%%%%%%%%%
\section{Molecular dynamics simulation of hard spheres}
\label{app:MDHS}
%%%%%%%%%%%%%%%%%%%%%%%%%%%%%%%%%%%%%%%%%%%%%%%%%

The numerical methods to compute the intermediate scattering function for a cubic  hard-sphere crystal with molecular dynamics are described.

\subsection{Hard-sphere dynamics}
\label{ssec:HS_dym}

We consider a hard-sphere system of $N$ identical spheres, each with mass $m$ and diameter $d$. The $i^{\rm th}$ sphere has a position ${\bf r}_i$ and a momentum ${\bf p}_i$. The spheres are within the simulation box of volume $V=L^3$, where $L$ is the length of the box edges, which are aligned with $x$, $y$, and $z$ axes. We apply periodic boundary conditions. Interactions between the spheres are mediated by a pair potential that is infinite if the interparticle distance $r_{ij}\equiv \Vert{\bf r}_i-{\bf r}_j\Vert$, computed with the minimum image convention~\cite{H97}, is less than $d$,  and vanishes otherwise. Consequently, particles do not overlap, and the Hamiltonian of the system is the conserved kinetic energy. The simulation is performed in the $(N,V,E)$ ensemble. The total momentum ${\bf P}=\sum_{i=1}^N{\bf p}_i$ is also conserved. The total energy is related to the temperature by $E = 3Nk_{\rm B} T/2$.

We use an event-driven algorithm~\cite{H97} for the simulation. The system evolves through free flights that are interrupted by instantaneous elastic collisions at times $\{t_c\}$.  When two particles collide, they exchange momentum. The change in momentum is given by $\Delta {\bf p}^{(c)}_{ij}=-({2}/{d^2})[{\bf r}^{(c)}_{ij}\cdot{\bf p}^{(c)}_{ij}(t_c-0)]{\bf r}^{(c)}_{ij}$, where ${\bf r}_{ij}\equiv{\bf r}_i-{\bf r}_j$ and ${\bf p}_{ij}\equiv({\bf p}_i-{\bf p}_j)/2$ are the canonically conjugate positions and momenta, respectively. Between collisions, the position of the $i^{\rm th}$ particle evolves as a free flight: ${\bf r}_i(t)={\bf r}_i(t_c)+{\bf p}_i(t_c+0)(t-t_c)/m$ with $t\in[t_c,t_{c+1})$. The list of  future collisions is found by solving the quadratic equations $[{\bf r}_{i_c}(t)-{\bf r}_k(t)]^2=d^2$ for $t  >t_c$, where $i_c$ labels the two particles colliding at time $t_c$ and $k$ represents all the other particles. 

In the simulation, the diameter and the mass of the spheres are set  to $1$.  Initial momenta of the particles are sampled from a Gaussian distribution, with the constraints that the total momentum ${\bf P}=0$ and that the temperature $k_{\rm B}T=1$. The algorithm generates trajectories of the hard spheres, providing their positions  and momenta at any time $t$ during the simulation. After a  transient evolution of time $t_{\rm transient}$, data is collected. We compute the ensemble averages of a quantity $X$ using $N_{\rm stat}$ trajectories, each with a length of $n_{\rm step}\Delta t$. Here,  $\Delta t$ is the time increment and $n_{\rm step}$ is the number of steps. The ensemble average is calculated as $\langle X(t_i)\rangle=N_{\rm stat}^{-1}\sum_{k=1}^{N_{\rm stat}}X(t^{k}_i)$, where $t^{k}_i=t_{\rm transient}+[(k-1)n_{\rm step}+{i-1}]\Delta t$, with $i\in\{1,...,n_{\rm step}\}$. Unless explicitly stated, we use $N_{\rm stat} = 2\times10^4$, $n_{\rm steps}=10^4$, and $\Delta t=0.01$ in the simulations.

The results are given in terms of dimensionless quantities, which are denoted by an asterisk.  The dimensionless particle density, positions, momenta, and time are defined as: $n_{0*} \equiv n_0d^3 = N_0d^3/V$, $ {\bf r}_{i*} \equiv   {\bf r}_{i}/d$, ${\bf p}_{i*} \equiv {\bf p}_{i}/ \sqrt{mk_{\rm B}T}$, and $t_* \equiv (t/d) \sqrt{k_{\rm B}T /m}$. The wave number, the frequency, and the correlation functions are expressed as 
 \begin{align}
q & = \frac{q_*}{d}\, , & \omega &= \frac{\omega_*}{d}\, \sqrt{\frac{k_{\rm B}T}{m}} \, ,& \frac{F({\bf q},t)}{S({\bf q})} &= \left[ \frac{F({\bf q},\omega)}{S({\bf q})} \right]_*  \, ,
\end{align}
in terms of their corresponding dimensionless quantities. 

\subsection{Simulation of a hard-sphere crystal with vacancies}

The hard-sphere model is  found in a fluid phase at densities below the freezing density $n_{\rm f*}=0.938 \pm 0.003$ and in a stable solid fcc phase for densities between the melting density  $n_{\rm m*}= 1.037 \pm 0.003$ and the close-packing density $n_{\rm cp*}= \sqrt{2}$~\cite{Sp97,Sp98,PBBCH19}. Between $n_{\rm f*}$ and $n_{\rm m*}$, the two phases coexist. The fcc lattice is composed of cubic cells, with $M$ cells along each spatial dimension. The edge length of each cell is $a$, so the total simulation box length is $L=aM$. The lattice sites within each cell are defined by four nodes at positions  ${\bf R}_j =x_j {\bf e}_x +y_j {\bf e}_y + z_j {\bf e}_z$ with $j\in\{1,2,3,4\}$ with $a^{-1}(x_j,y_j,z_j)\in\{(\frac{1}{4},\frac{1}{4},\frac{1}{4}),(\frac{3}{4},\frac{3}{4},\frac{1}{4}),(\frac{3}{4},\frac{1}{4},\frac{3}{4}),(\frac{1}{4},\frac{3}{4},\frac{3}{4})\}$~\cite{AM76}. This gives a total of $N_0=4M^3$  lattice sites and an equilibrium density of  $n_0=4/a^3=N_0/V$. Initially, $N$ particles are  placed  on these lattice nodes. For a crystal with vacancies,  a number $N_{\rm v}$ of lattice sites are left unoccupied, such that $N+N_{\rm v}=N_0$. If the crystal is perfect, $N_{\rm v}=0$, and all the sites of the lattice are occupied by exactly one particle. During the simulation  in the crystalline phase, the numbers $N_0$, $N$, and $N_{\rm v}$ remain constant. 

At any given time during the simulation,  vacancies can be located  by a two-step process. First, each particle is matched to its closest lattice site. For a particle $i$, the closest lattice site is given by  ${\rm min}_{j}[{\bf r}_i(t)-{\bf R}_j(t)]^2$, where ${\bf R}_j(t)$ is the position of the $j^{\rm th}$ lattice node at time $t$. Vacancies are then identified as the remaining unmatched sites. Due to the conservation of total momentum in the simulation, the center-of-mass position is also conserved. Consequently, each time a particle jumps to a nearest-neighbor site at a distance $a/2$, the lattice drifts by a distance $a/(2N)$ in the opposite direction~\cite{vdMDF17}. The position of a site at time $t$ is thus ${\bf R}_j(t)={\bf R}_j(0)+\Delta {\bf R}(t)$, where $\Delta {\bf R}(t)$ is the drift of the lattice. This drift is determined by the number of times the vacancies jump in a given direction, multiplied by the  lattice drift $a/(2N)$.  The vacancies are always located with respect to the position of the lattice at time $t$.  Note that locating the vacancies is not needed to compute the intermediate scattering function. However, their positions are used to compute $S_{\rm nv}({\bf q})$ that enters the analytical expression~\eqref{eq:DSF_Full}.
 
The assumption of conservation of the number of vacancies is justified by the absence of any event where a vacancy-interstitial pair is generated in the numerical simulations. The rate of such pair creation is thus negligible under the conditions considered.

\subsection{Spectral functions from the hard-sphere dynamics}

From the  positions ${\bf r}_i(t)$ of the particles, the density is defined by $\hat{n}({\bf r},t)\equiv\sum_{i=1}^N\delta[{\bf r}-{\bf r}_i(t)]$ and its Fourier transform by 
\begin{align}
\label{eq:n(q)}
\hat{n}({\bf q},t)=\int_V\text{d}{\bf r}\;{\rm e}^{{\rm i}{\bf q}\cdot {\bf r}}\; \hat n({\bf r},t) =\sum_{i=1}^N {\rm e}^{{\rm i} {\bf q}\cdot{\bf r}_i(t)} \, .
\end{align}
Due to the periodic boundary conditions, the density has the periodicity $\hat{n}({\bf r},t)=\hat{n}({\bf r}+{\bf L},t)$ with ${\bf L}\equiv L\left(m_x{\bf e}_x+m_y{\bf e}_y+m_z{\bf e}_z \right)$ and $(m_x,m_y,m_z)\in {\mathbb Z}^3$. The wave vector ${\bf q}$ of the Fourier modes defined in this domain is discrete, and is given by ${\bf q} = (2\pi/L)\left(n_x{\bf e}_x+n_y{\bf e}_y+n_z{\bf e}_z \right)$ with $(n_x,n_y,n_z)\in {\mathbb Z}^3$. The special directions of the wave vector correspond to $n_y=n_z=0$ for $[100]$, $n_x=n_y$, $n_z=0$ for $[110]$ and $n_x=n_y=n_z$ for $[111]$.

In the simulation, we compute the intermediate scattering function, which is given as
\begin{align}
F({\bf q},t)&=\frac{1}{N}\langle   \delta\hat{n}({\bf q},t)    \delta\hat{n}^*({\bf q},0)  \rangle =\frac{1}{N}\langle   \hat{n}({\bf q},t)    \hat{n}^*({\bf q},0)  \rangle - \frac{1}{N} \left|\langle n({\bf q})\rangle\right|^2,\label{eq:ISF_MDHS}
\end{align}
using that $ \delta\hat{n}({\bf q},t)   =\hat{n}({\bf q},t)-\langle \hat n({\bf q})\rangle$. Unlike for a perfect crystal~\cite{MG24b}, in the presence of vacancies, the  term $\left|\langle n({\bf q})\rangle\right|^2$ is non-zero for $q$ in the hydrodynamic regime. However, this contribution is of the order of ${\cal O}(y)$.

\end{document}